\documentclass[journal]{IEEEtran}
\usepackage{color}
\usepackage{bbding}
\usepackage{pifont}
\usepackage{wasysym}
\usepackage{amssymb}
\usepackage{amsthm}
\usepackage{booktabs}
\usepackage{multirow} 
\usepackage{setspace}
\usepackage{epsfig}
\usepackage{graphicx}
\usepackage{subfigure}
\usepackage{algorithm}
\usepackage{algpseudocode}
\usepackage{url}

\usepackage{array}
\newcolumntype{L}[1]{>{\raggedright\let\newline\\\arraybackslash\hspace{0pt}}m{#1}}
\newcolumntype{C}[1]{>{\centering\let\newline\\\arraybackslash\hspace{0pt}}m{#1}}
\newcolumntype{R}[1]{>{\raggedleft\let\newline\\\arraybackslash\hspace{0pt}}m{#1}}

\usepackage{xcolor}

\ifCLASSINFOpdf
\else
\fi
\begin{document}
	%
	% paper title
	% Titles are generally capitalized except for words such as a, an, and, as,
	% at, but, by, for, in, nor, of, on, or, the, to and up, which are usually
	% not capitalized unless they are the first or last word of the title.
	% Linebreaks \\ can be used within to get better formatting as desired.
	% Do not put math or special symbols in the title.
	\title{From IC Layout to Die Photo: A CNN-Based Data-Driven Approach}
	%
	%
	% author names and IEEE memberships
	% note positions of commas and nonbreaking spaces ( ~ ) LaTeX will not break
	% a structure at a ~ so this keeps an author's name from being broken across
	% two lines.
	% use \thanks{} to gain access to the first footnote area
	% a separate \thanks must be used for each paragraph as LaTeX2e's \thanks
	% was not built to handle multiple paragraphs
	%
	
	\author{Hao-Chiang~Shao,~\IEEEmembership{Member,~IEEE}, Chao-Yi~Peng, Jun-Rei~Wu, Chia-Wen~Lin,~\IEEEmembership{Fellow,~IEEE}, Shao-Yun~Fang,~\IEEEmembership{Member,~IEEE}, 
	Pin-Yian~Tsai, and~Yan-Hsiu~Liu% <-this % stops a space
	\thanks{Manuscript received on February 11, 2020; revised May 25, 2020; accepted July 16, 2020. Date of publication Month Date, 2020; date of current version Month Date, 2020. This work was supported in part by the Ministry of Science and Technology, Taiwan, under Grants MOST 108-2634-F-007-009, and in part by United Microelectronics Corporation. The associate editor coordinating the review of this manuscript and approving it for publication was Dr. Laleh Behjat.}
	\thanks{Hao-Chiang Shao is with the Department of Statistics and Information Science, Fu Jen Catholic University, Taiwan. (e-mail:shao.haochiang@gmail.com)}
	\thanks{Chao-Yi Peng and Jun-Rei~Wu were with the Department of Electrical Engineering, National Tsing Hua University, Hsinchu, Taiwan.}		
	\thanks{Chia-Wen Lin (corresponding author) is with the Department of Electrical Engineering and the Institute of Communications Engineering, National Tsing Hua University, Hsinchu, Taiwan. (e-mail: cwlin@ee.nthu.edu.tw)}
	\thanks{Shao-Yun Fang is with the Department of Electrical Engineering, National Taiwan University of Science and Technology, Taipei, Taiwan. (e-mail: syfang@mail.ntust.edu.tw)}
	\thanks{Pin-Yian Tsai and Yan-Hsiu Liu are with United Microelectronics Corporation, Hsinchu, Taiwan. (e-mail: \{pin\_yian\_tsai; cecil\_liu\}@umc.com)}
	\thanks{Color versions of one or more of the figures in this paper are available online at http://ieeexplore.ieee.org.}
	}
	
	% note the % following the last \IEEEmembership and also \thanks - 
	% these prevent an unwanted space from occurring between the last author name
	% and the end of the author line. i.e., if you had this:
	% 
	% \author{....lastname \thanks{...} \thanks{...} }
	%                     ^------------^------------^----Do not want these spaces!
	%
	% a space would be appended to the last name and could cause every name on that
	% line to be shifted left slightly. This is one of those "LaTeX things". For
	% instance, "\textbf{A} \textbf{B}" will typeset as "A B" not "AB". To get
	% "AB" then you have to do: "\textbf{A}\textbf{B}"
	% \thanks is no different in this regard, so shield the last } of each \thanks
	% that ends a line with a % and do not let a space in before the next \thanks.
	% Spaces after \IEEEmembership other than the last one are OK (and needed) as
	% you are supposed to have spaces between the names. For what it is worth,
	% this is a minor point as most people would not even notice if the said evil
	% space somehow managed to creep in.

	% The paper headers
	\markboth{IEEE Transactions on Computer-Aided Design of Integrated Circuits and Systems,~Vol.~x, No.~x, Month~2020}%
	{Shell \MakeLowercase{\textit{et al.}}: Bare Demo of IEEEtran.cls for IEEE Journals}
	% The only time the second header will appear is for the odd numbered pages
	% after the title page when using the twoside option.
	% 
	% *** Note that you probably will NOT want to include the author's ***
	% *** name in the headers of peer review papers.                   ***
	% You can use \ifCLASSOPTIONpeerreview for conditional compilation here if
	% you desire.

	% If you want to put a publisher's ID mark on the page you can do it like
	% this:
	%\IEEEpubid{0000--0000/00\$00.00~\copyright~2015 IEEE}
	% Remember, if you use this you must call \IEEEpubidadjcol in the second
	% column for its text to clear the IEEEpubid mark.

	% use for special paper notices
	%\IEEEspecialpapernotice{(Invited Paper)}

	% make the title area
	\maketitle
	
	% As a general rule, do not put math, special symbols or citations
	% in the abstract or keywords.
	\begin{abstract}

%	Since IC fabrication is costly and time-consuming, it is highly desirable to develop virtual metrology tools that can predict the properties of a wafer based on fabrication configurations without performing physical measurements on a fabricated IC. 
%===
%The existing lithography simulation program is time-consuming due to its highly computational complexity.
We propose  a deep learning-based data-driven framework consisting of two convolutional neural networks: i) LithoNet that predicts the shape deformations on a circuit due to IC fabrication, and ii) OPCNet that suggests IC layout corrections to compensate for such shape deformations. By learning the shape correspondences between  pairs of  layout design patterns and their scanning electron microscope (SEM) images of the product wafer thereof, given an IC layout pattern, LithoNet can mimic the fabrication process to predict its fabricated circuit shape.
%In addition, we formulate the wafer fabrication parameters as a latent vector to model the parametric product variance that can be inspected on the SEM images.
Furthermore, LithoNet can take the wafer fabrication parameters as a latent vector to model the parametric product variations that can be inspected on SEM images. 
%the wafer fabrication parameters are considered as a latent vector in our LithoNet to model the parameteric product variance that can be inspected on the SEM images.
% Consequently, the predicted image of fabricated IC is modeled as a function of fabrication parameters and input layout by LithoNet. 
%
%
%
%Moreover, existing lithography simulation algorithms used for suggesting a correction on a lithographic photomask according to a given layout by checking whether the shape of a fabricated IC circuitry matches exactly the layout design, is computationally very expensive. Thus, we propose an OPCNet generator, cooperating with a pre-trained LithoNet, to mimic the OPC (optical proximity correction) procedure used to correct the layout design and generate a photomask. 
Besides, traditional optical proximity correction (OPC) methods used to suggest a correction on a lithographic photomask is computationally expensive. Our proposed OPCNet mimics the OPC procedure and efficiently generates a corrected photomask by collaborating with LithoNet to examine if the shape of a fabricated circuit optimally matches its original layout design. 
As a result, the proposed LithoNet-OPCNet framework can not only predict the shape of a fabricated IC from its layout pattern, but also suggests a layout correction according to the consistency between the predicted shape and the given layout. Experimental results with several benchmark layout patterns demonstrate the effectiveness of the proposed method.
%
%Moreover, because the wafer image that go through fabrication process will not as same as desired original layout image because of exposures, chemical reactions and physical properties. Thus we use an additional generator followed by the pre-trained LithoNet and forcing the generator to learn how to revise the original layout and generate a mask by utilizing the proposed input-output consistency loss. We achieve unsupervised mask optimization by limiting the output simulation result of the mask and the original layout as same as possible.
%
%Through this design, the synthesized SEM-styled image can be modeled as a function of fabrication parameters. 
%We evaluate our LithoNet-OPCNet system using various benchmark layout patterns, and our experimental results demonstrate the effectiveness of it.
	\end{abstract}
	
	% Note that keywords are not normally used for peerreview papers.

	% For peer review papers, you can put extra information on the cover
	% page as needed:
	% \ifCLASSOPTIONpeerreview
	% \begin{center} \bfseries EDICS Category: 3-BBND \end{center}
	% \fi
	%
	% For peerreview papers, this IEEEtran command inserts a page break and
	% creates the second title. It will be ignored for other modes.
	\IEEEpeerreviewmaketitle

%------------------------------------------------------------------------	
	\section{Introduction}
	\label{sec:intro}
	%It is essential in semiconductor manufacturing to develop a virtual metrology method that can predict the properties of a wafer based on fabrication configurations, such as parameters and layout design, without performing a physical measurement of the wafer produced after the whole costly fabrication process.  

% [2019.6.2]
% OPCNet要再斟酌一下, 雖然說我們知道 OPC的目的是修正layout mask
% 但是在系統的設計上, 
% 其實是要強調 lithonet + opcnet, 可以構成一個自我估測的 loop,
% 正向可以評估 sem,
% 反向可以修正 layout
%
% 反向的修正, 雖然OPC觀念上是 layout-to-mask,
% 但我們的做法是從 sem 來反推 layout correction.
%
%
% 先預備定調成 
%  LithoNet:  Layout-to-SEM
%  OPCNet: SEM-to-Layout(Mask
%
% 

\IEEEPARstart{A}{fter} IC circuit design and layout, it typically takes two to three months to fabricate a 12-inch IC wafer, involving a multi-step sequence of photolithographic and chemical processing steps. Among these steps, a lithography process is used to transfer an IC layout pattern from a photomask to a photosensitive chemical photoresist on the substrate, followed by an etch process that chemically removes parts of a polysilicon or metal layer, uncovered by the etching mask, from the wafer surface. Because it is hard to control the exposure conditions and the chemical reactions involved in all fabrication steps, the two processes together lead to nonlinear shape distortion of a designed IC pattern, which is usually too complicated to model. This fact urges the need for \textit{mask optimization}, a procedure that computes an optimized photomask  to make the shape of the fabricated IC wafer optimally consistent with its source layout design.
The inevitable shape deformations on a fabricated IC due to the imperfect lithography and etch processes often cause IC defects (e.g., thin wires or broken wires) if an IC circuit layout is not appropriately designed, especially on the first few metal layers. Nevertheless, in most cases we still cannot identify such IC defects due to inappropriate IC circuit layout until capturing and analyzing the scanning electron microscope (SEM) images of metal layers after the wafer fabrication process, making the circuit verification very costly and time-consuming. 
%
%
%It is therefore desirable to predict the corresponding SEM  image of metal lines from an IC layout and assess an optimized photomask for etching in a pre-simulation process. 
It is therefore desirable to develop pre-simulation tools, including i) a lithography simulation method for predicting the shapes of fabricated metal lines based on a given IC layout along with IC fabrication parameters, and ii) a mask optimization strategy for predicting the best mask to compensate for the shape distortions caused by the lithography and etch processes.
% an IC layout and assess the IC layout quality in a pre-simulation process.

%\noindent \underline{\textbf{Boldface: rewrite}}

%\textbf{Besides, after a layout pattern go through a fabrication process, the etching result usually not meet the original desire layout due to different exposures, chemical reactions and physical properties. Thus, it is essential to modify the original IC layout to generate a mask, trying to make the etching results of the mask meeting the original IC layout, so-called mask optimization.}

\begin{figure*}[!t]
\centering
\includegraphics[width=0.75\textwidth,keepaspectratio=true]{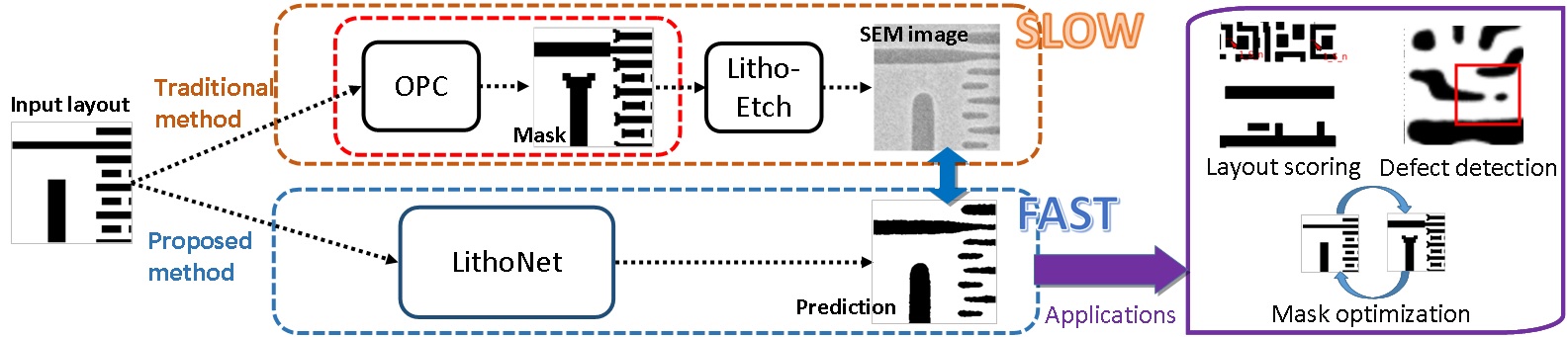} %\par \\ [-0.3cm]
\vspace{-0.3cm}
\caption{Relationship among OPC simulation, circuit verification on an SEM image, and our method. 
The OPC step, highlighted by the red dashed lines, suggests modifications of a layout mask so that the fabricated IC could have nearly the same shape as the original layout pattern. The proposed LithoNet and its applications are highlighted by purple contours.}
\label{fig:metrology}
\end{figure*}

As for lithography simulation, there are two categories of conventional approaches: physics-level rigorous simulation and compact model-based simulation \cite{watanabe2017accurate,ye2019lithogan}. Rigorous simulation methods simulate physical effects of materials to accurately predict a fabricated circuit and thus are very time-consuming \cite{taflove2005computational,lucas1996efficient}. On the contrary, a compact model-based simulation method follows loosely physical phenomena to obtain a faster computational speed by exploiting complicated, parameter-dependent, non-linear functions. 
Different from traditional methods, we aim at developing a convolutional neural network (CNN) based approach, which learns the parametric model of physical and chemical phenomena of a fabrication process directly from a training dataset containing pairs of IC layouts and their corresponding SEM images. Based on the learned CNN model, we can predict a fabricated circuit shape more accurately and efficiently than traditional methods. 

Moreover, fab-engineers usually optimize a mask pattern by iteratively modifying a layout design based on its lithography simulations. However, rule-based lithography simulations resort to linear combinations of optical computations derived from several similar yet not identical historical fab-models. %This fact may make conventional mask optimization methods unreliable against new layout patterns. 
%============Revision, R1Q1: old content==========
%The simulation reliability largely relies on a rich amount of historical fabrication data in the database, which is, however, very costly because ground-truth fab-models need to be gathered by fabricating a layout pattern with all possible configurations. 
%=====R1Q1, new content below=====
The simulation reliability largely relies on a rich amount of costly historical fabrication data because ground-truth fab-models need to be gathered by fabricating a layout pattern with all process variations exhaustively. Nevertheless, fab-plants do not typically build models with exhaustive data but, instead, select nominal plus some relatively small number of specific process-window conditions over a limited number of test structures and then build models based on that. This fact may make current standard unreliable for new layout design patterns.

%Fig. \ref{fig:metrology} depicts the relationship among the IC fabrication process, lithography simulator, and mask optimizer. %,. where the 
The relationship among the IC fabrication process, lithography simulator, and mask optimizer is depicted in Fig. \ref{fig:metrology}, where the OPC (optical proximity correction) block is a standard approach to photomask correction for compensating for the shape distortions due to diffraction or process effects as well as guaranteeing the printability of a layout pattern, especially at the corners of the process window \cite{otto1994automated,hsu2001optical}. As shown in the red dashed rectangles in Fig. \ref{fig:metrology}, the mask used in the fabrication process is a modified version of a source layout design, aiming to compensate for possible ``shrinkages" in line shapes due to the fabrication  to mitigate the deviation of a fabricated IC circuitry from its layout design.
However, traditional OPC methods have two primary drawbacks. First, they run simulations based on those rules and patterns already known; thus, an OPC correction may be unreliable if an unseen layout design is given. Second, not only is the OPC correction computationally expensive, but also the OPC contour simulation is a time-consuming trial-and-error routine that is iterated until no irregularity can be found in the OPC estimation result. 
%A layout needs to be modified according to its OPC test again and again until no irregularity can be found on the OPC estimation. 
%
%
%==以下R1Q2刪除==
%Due to its high complexity, the OPC simulation is usually performed on a limited number of regions of interest (ROIs) rather than on the whole layout design to reduce computation. 
%==以下R1Q2新增==
Both the OPC correction and OPC contour simulation are computationally expensive.
Take the \textit{ICWB} software (\textit{IC WorkBench}) developed by Synopsys \cite{Synopsys} for example. ICWB takes, on average, about 34 seconds to run a contour simulation on a $4 \times 1.7 \mu m^2$ layout patch with an Intel Xeon E5-2670 CPU and 128GB RAM. It will cost around 4 days to run one OPC contour simulation on a $400 \times 170 \mu m^2$ layout design, and such computational cost makes a complete OPC contour simulation procedure impractical. It is therefore highly desirable to develop an efficient photomask optimization scheme.
%As for the photomask optimization issue, because we don’t have the ground truth optimal mask, we utilize the already trained lithography simulator, LithoNet, as an auxiliary network to train an additional mask generator that can generate modified masks. With the proposed input-output consistency loss to achieve unsupervised mask optimization.

Recent progress on image-to-image translation techniques makes them suitable to tackle the lithography simulation (i.e., Layout-to-SEM) and photomask optimization (i.e., SEM-to-Layout) problems mentioned above. 
%\textbf{
However, these two issues are more complicated than general image-to-image translation problems. Take Layout-to-SEM prediction for example.
%}
First of all, the domain of IC layout images and that of SEM images are heterogeneous. An IC layout is a purely man-made blueprint with only lines and rectangles on it, and hence it is noise-free and artifact-free. On the contrary, an SEM image is formed from the intensity of detected signal from raster-scanning the IC surface with a focused electron beam. Besides the continuous shape distortions introduced by the lithography and etching processes, the SEM imaging process itself also suffers from several kinds of interference (e.g., scan-line noise and shading). This fact leads SEM images to a significantly different domain from the layout-image domain. Hence, this issue is essentially a cross-domain image matching and translation problem. 
Second, in order to predict the corresponding SEM image from an IC layout, our solution must be capable of finding the shape correspondence between these two domains of images. %, as has not been a primary concern in general image-to-image translation techniques. 
This fact raises an unsupervised cross-domain image matching issue, which usually has not been concerned in general image-to-image translation techniques. Thus it requires a more sophisticated solution, as the concerns stated in \cite{aberman2018neural,zhou2016learning}. Third, for the mask optimization problem, it is very costly to collect a comprehensive set of reference OPC-corrected photomasks, making the training of a photomask optimization network infeasible.

\begin{figure}
    %\centering
    \includegraphics[width=0.48\textwidth,keepaspectratio=true]{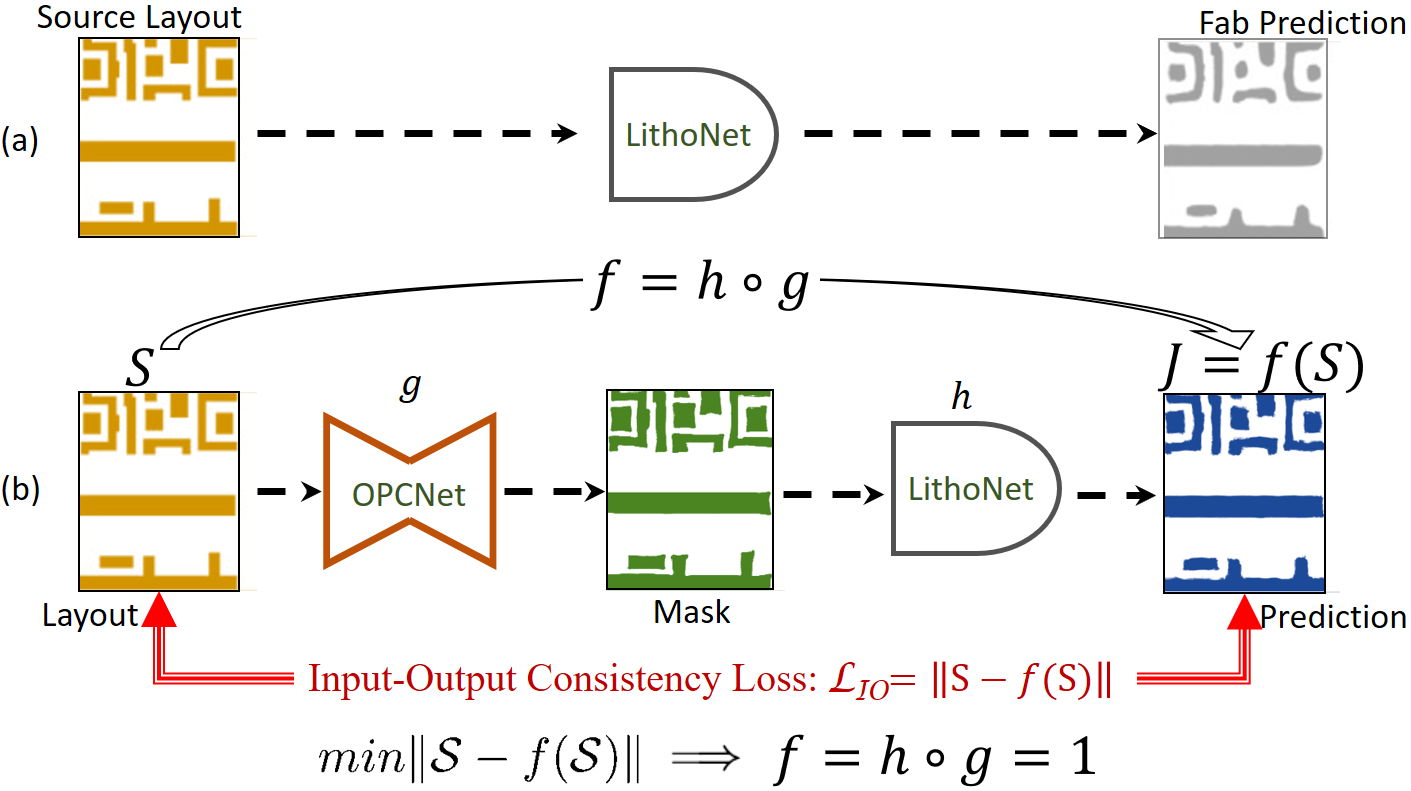} 
    %Framework_new.jpg}%{figures/framework.PNG}
    \caption{Two scenarios utilizing the proposed LithoNet and OPCNet: (a) A stand-alone LithoNet, and (b) A cascaded LithoNet-OPCNet network.}
    \label{fig:framework}
\end{figure}

To address the above problems, as shown in Fig. \ref{fig:framework}, we propose a fully data-driven framework involving two CNNs,  LithoNet and OPCNet, functionally complementary to each other.
 In short, LithoNet is a cross-domain simulator of the lithography and etch processes in IC fabrication, and OPCNet is a self-supervised mask optimization CNN using the prediction results of LithoNet as supervision for the purpose of OPC.
%To overcome this difficulty, we design a framework in which we utilize the aforementioned CNN model as an auxiliary module to train a mask-generator. Supposing that our CNN-based lithography simulation network is well-trained, we aim to this idea makes our mask-generator groundtruth-free and self-supervised. 
The proposed LithoNet-OPCNet network serves two purposes, each requiring a specific training dataset. 
First, when LithoNet is used stand-alone as shown in Fig. \ref{fig:framework}(a), it aims at image-to-image contour prediction. Because we focus on the Layout-to-SEM (or Mask-to-SEM) contour prediction problem, we train LithoNet on (layout, SEM) data pairs. Then, during the inference stage, given a layout design, LithoNet predict i) a deformation map and ii) an SEM prediction. Both the deformation map and SEM prediction can be used for layout risk assessment. Note that LithoNet is an image-to-image contour predictor and thus can be trained on different kinds of paired images for different purposes. For example, if we need to build a model for mask-to-SEM prediction, we have to train LithoNet on (mask, SEM) data pairs. 

Second, as shown in Fig. \ref{fig:framework}(b), when OPCNet and LithoNet are cascaded, the LithoNet-OPCNet network forms a system for mask optimization aiming at minimizing the discrepancy between a source layout and its SEM contour predicted by LithoNet. The design concept is to construct a two-stage system, where the first stage performs layout-to-X prediction by OPCNet, where X denotes the OPC-corrected mask, and the second stage performs X-to-SEM prediction by LithoNet. Then, by enforcing the SEM prediction to be shape-consistent with the target layout (i.e., the whole OPCNet-LithoNet network behaves as an identity transform), OPCNet and LithoNet act as if they were inverse functions of each other mathematically. As a result, the LithoNet-OPCNet network can be used to find an OPC-optimized mask X.

This paper has four primary contributions: 
\begin{itemize}
\item To the best of our knowledge, we are the first to formulate the Layout-to-SEM deformation prediction problem as a cross-domain image correspondence problem, and we propose a two-step CNN-based framework to address it. 
\item Our LithoNet-OPCNet system is computationally much more efficient than the typical optical-based contour simulation schemes, while achieving comparable prediction accuracy. Therefore, since our method is fully data-driven, it could enable IC fabrication plants to run a full, large-scale screening on new IC layout designs. 
%Note that the standard OPC approaches rely on sophisticated design rules and design patterns already in the database, and thus it can only examine a limited amount of areas each time. 
Note that an OPC model is typically built for a particular process condition and operates according to interpolation. Hence, if the process condition changes for a given process node, whether the input layout is completely new or not, the same OPC model may not provide a reliable prediction. 

\item The proposed LithoNet is parameterized with fabrication settings. Hence, it can also predict results under different fabrication conditions so as to assist fabrication plants to find the best suitable working intervals of parameters and thus be beneficial for yield-rate improvement. 
\item The proposed OPCNet overcomes the difficulty in lack of ground-truth mask patterns. With the aid of a novel training objective function called \textit{I/O-consistency} loss, the proposed OPCNet can well simulate the mask optimization process in collaboration with LithoNet.
\end{itemize}

The remainder of this paper is organized as follows. We review related literature in Section \ref{sec:related}. The proposed LithoNet and OPCNet are detailed in Sections \ref{sec:lithonet} and \ref{sec:opcnet}, respectively. Section \ref{sec:experiments} demonstrates and discusses our experimental results. Finally, we draw our conclusion in Section \ref{sec:conclusion}.

	\section{Related Work}
	\label{sec:related}
	%\subsection{Virtual Metrology}

\subsection{Virtual Metrology}
\label{subsec:previousVM}

In IC fabrication, virtual metrology (VM) refers to the methods for predicting wafer properties based on fabrication parameters and sensor data from equipment without performing physical measurements on the product wafer produced by a whole, costly fabrication process \cite{hung2007novel}. 
Since VM techniques can significantly reduce the cost of IC fabrication, various kinds of VM methods have been proposed for fabrication quality assessment.
%Virtual metrology methods were application-oriented designs. 
For example, % as for the prediction of average Silicon Nitride cap layer thickness,
%for the Plasma Enhanced Chemical Vapor Deposition (PECVD) dual-layer metal passivation stack process regression-based VM methods were %designed for this purpose developed as surveyed in \cite{purwins2014regression}. Specifically,
Susto et al. exploited the knowledge collected in the process steps to improve the accuracy of  VM prediction via a multi-step strategy \cite{susto2015multi}. 
Besides, the demand of VM methods has also triggered the development of theoretical techniques. The method in \cite{poonawala2007mask}, for instance, models OPC mask correction as an inverse problem of optical microlithography. Optical lithography is a process used for transferring binary circuit patterns onto silicon wafers, and related discussions about lithography techniques can be found in \cite{pan2013design}. Recently, people have been attempting to integrate machine learning methods with IC implementation and VM \cite{watanabe2017accurate,ye2019lithogan,kahng2018reducing,GANOPC2018yang,yu2019deep}. Specifically, Yang et al. proposed in \cite{GANOPC2018yang} a generative adversarial network (GAN) \cite{goodfellow2014generative} based  inverse method to estimate the optimal mask used in the fabrication process from an OPC simulation result. 
However, the design in \cite{GANOPC2018yang} aims only at the OPC-to-Layout problem, which operates in an opposite direction of our Layout-to-SEM prediction. 
%\textbf{
Therefore, to the best of our knowledge, there is no existing technique focusing simultaneously on both Layout-to-SEM (lithography simulation) and SEM-to-Layout (mask optimization) image translation problems. We deem that a hybrid method of image-to-image translation or feature mapping techniques could compose a straightforward solution to these two prediction problems.

\subsection{Lithography Simulation}
\label{subsec:previousLitho}

%\textbf{\underline{(Rewrite)}}
%\textbf{Recently, some people began to do lithography simulation base on machine learning methods. Watanabe et al. [17] proposed a convolutional neural network (CNN) model to achieve accurate simulation to overcome the explicit feature extraction. W. Ye et al. [15] map the input mask patterns directly to the output resist patterns by an end-to-end lithography modeling framework based on a generative adversarial network (GAN). They design a dual learning network that predicts the resist shape using cGAN  model and predicts resist center using a CNN model. Our idea is as same as [15], formulating the Layout-to-SEM prediction problem as an image-to-image translation task.}

%\textbf{
%Lithography simulation methods have been developed based on machine learning techniques recently. 
Recently, there have been a few machine learning-based lithography simulation methods.
For instance, Watanabe et al. proposed a fast and accurate lithography simulation by determining an appropriate model function via CNN \cite{watanabe2017accurate}, and Ye et al. developed a GAN-based end-to-end lithography modeling framework, named LithoGAN, to map directly the input mask pattern to the output resist pattern \cite{ye2019lithogan}. 
Specifically, LithoGAN models the shape of the resist pattern based on a conditional GAN (cGAN) model and predicts the center location of the resist pattern via a CNN model. 
LithoGAN has a dual learning framework, and similarly our LithoNet also adopts a dual learning framework.
%}

\begin{figure*}[!t]
%\begin{tabular}{ p{180pt}}
\centering
\includegraphics[width=0.95\textwidth,keepaspectratio=true]{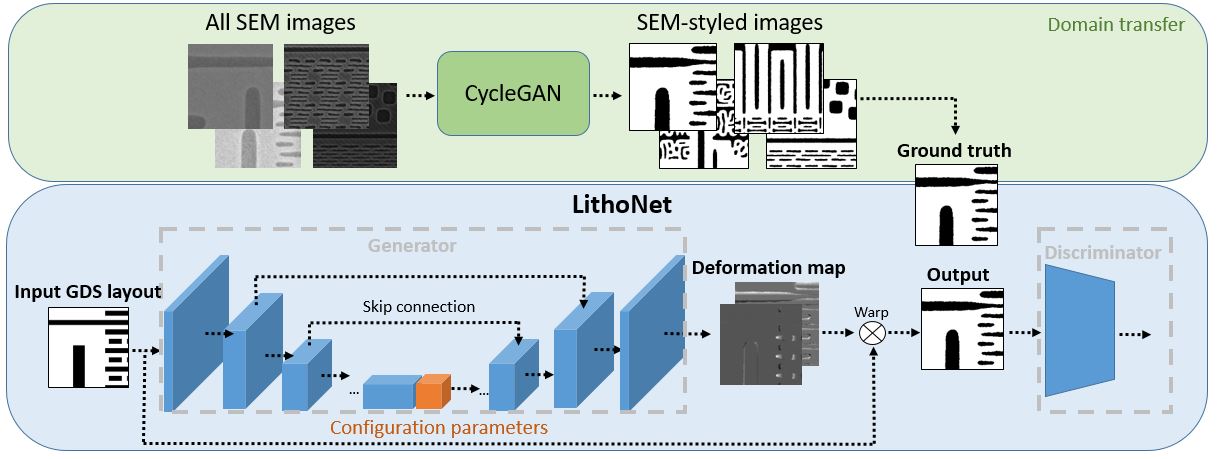}% \par \\ [-0.1cm]
%\end{tabular}
\caption{Block diagram of the proposed two-step framework for cross-domain image-to-image translation. The upper step adopts CycleGAN to transfer the training SEM images to obtain ground-truth labels. LithoNet then estimates the deformation maps between input layout patterns and their corresponding labels.}
\label{fig:LithoNet}
\end{figure*}

%Moreover, as will be detailed in Section \ref{sec:lithonet}, in the LithoNet design, we formulate the Layout-to-SEM prediction as a cross-domain image-to-image translation problem. 
As will be detailed in Section \ref{sec:lithonet}, we formulate the Layout-to-SEM prediction as a cross-domain image-to-image translation problem in the LithoNet design.
%Recently, several image-to-image translation methods were proposed. These works can be divided into two groups: 
Recent image-to-image translation methods can be divided into two groups. One requires training image pairs, e.g., \cite{isola2017image,wang2018high}, and the other supports training on unpaired data, e.g., \cite{liu2017unsupervised}.
The method in \cite{liu2017unsupervised}, based on GANs \cite{goodfellow2014generative} and VAEs \cite{kingma2013auto}, was designed for unsupervised image-to-image translation tasks, which could be considered as a conditional image generation model. 
Besides, Pix2pix \cite{isola2017image} consists of a Unet-like generator and a PatchGAN discriminator. Pix2pix uses the PatchGAN discriminator to model high-frequencies by classifying if each patch in an image is real or fake. Therefore, it can be adopted in various applications, such as translating a cartoon map to a satellite image and translating a sketch to a natural image, and has become a benchmark in this field. Pix2pix was further enhanced in \cite{wang2018high} by taking advantage of a course-to-fine generator, a multi-scale discriminator, and an adversarial learning objective function so as to generate high-resolution photo-realistic images. 
 
%,taigman2016unsupervised}

%\textbf{
%The existing image-to-image translation methods may not be so appropriate for the Layout-to-SEM image translation for VM purpose because the shape correspondence or the deformation field is the key information for predicting whether IC line widths are narrower or wider than a designed layout. 
%The deformation field or the shape correspondence between layout and SEM images are thus quite essential for this purpose. 
However, none of the above methods addresses the shape correspondence and the deformation field between two different domains of images, and neither do other representative image-to-image translation methods, such as CycleGAN \cite{zhu2017unpaired}, DualGAN \cite{yi2017dualgan}, and \cite{liu2017unsupervised,bousmalis2017unsupervised,huang2018multimodal}. 
Because characterizing the deviations of metal lines on a product IC based on the source layout is a critical point in the IC industry, traditional image-to-image translation methods, which lack a mechanism for precisely estimating a deformation field or the shape correspondence between the layout and SEM images, are not applicable to Layout-to-SEM image translation.
%may not be so appropriate for the Layout-to-SEM image translation for VM purpose because the shape correspondence or the deformation field is the key information for predicting whether IC line widths are narrower or wider than a designed layout.
%}
To serve the above purpose, the proposed LithoNet model   performs cross-domain image-to-image translation via learning the shape correspondence between paired training images so as to output a predicted deformation map for further VM applications.

\subsection{Mask Optimization}
\label{subsec:previousMask}

%\textbf{There are some machine learning or deep learning based methods to do mask optimization like [14, 16]. Yang et al. [14] proposed a generative adversarial network (GAN) [23] based method to estimate the optimal mask used in the fabrication process from an OPC simulation result. To facilitate the training process and ensure better convergence, they also propose a pre-training procedure that jointly trains the neural network with inverse lithography technique (ILT) [29]. After model converge, the generated quasi-optimal masks are expected to be good initial for further ILT operation. Besides, Yu et al. [16] simultaneously perform sub-resolution assist feature (SRAF) [30] insertion and edge-based OPC base on a deep learning framework. However, both of [14] and [16] need a OPC ground truth mask dataset to train their network, but it is time-consuming to collect the optimal ground truth mask. Thus we proposed an unsupervised learning method by utilizing a new objective function called input-output consistency loss.}

%\textbf{
There also exist machine learning-based mask optimization approaches. Notably, GAN-OPC proposed in~\cite{GANOPC2018yang} takes source layout patterns and their reference OPC photomasks as training inputs and accordingly, for an input layout design, predicts a corrected photomask that minimizes the deviation on the (simulated) fabricated circuit shape from its original design. In order to facilitate the training process and guarantee convergence, GAN-OPC involves a pre-train procedure that trains jointly the neural network and the inverse lithography technique (ILT) \cite{gao2014mosaic}. After GAN-OPC converges, the obtained quasi-optimal photomask is further used as a reasonable initial estimate for further ILT operation. In contrast, Yu et al. \cite{yu2019deep} proposed a DNN framework to simultaneously perform  sub-resolution assist feature (SRAF) \cite{gabor2002subresolution} and edge-based OPC. However, the two methods require a collection of photomask images, such as those suggested by OPC or historical data gathered during the actual fabrication process, as the ground-truth dataset for training. Because it is expensive and time-consuming to collect qualified mask images, the cardinality of the training dataset forms a performance bottleneck of these methods. To eliminate such a bottleneck, we propose the OPCNet model for mask optimization, powered by LithoNet. Because OPCNet and LithoNet are the inverse function to each other, OPCNet can be trained directly on the SEM-styled images predicted by LithoNet without the need for using expensive photomask images, as will be elaborated later.
	\section{LithoNet: A CNN-Based Lithography Simulator}
	\label{sec:lithonet}
	%You must include your signed IEEE copyright release form when you submit your finished paper. We MUST have this form before your paper can be published in the proceedings.

%預計的包裝方式是:
% 兩個 domain 無法直接比較,也無法直接transfer,
% 所以採取折衷的策略, 將兩個domain的資料,
% 轉換到中間的某個媒介(intermediate reference domain).
%
% 而這個中間媒介的特色則是: 
% (1) 具有 SEM 的 shape contour  和 
% (2) 具有 layout 的 binary image特質
%
% 也就是說, 引入atlas, registration/warping的reference domain的概念
% 而這樣的中繼domain,也有望支援未來可能發展的virtual metrology
% 這要寫進intro
%
%
% 這樣會比較好包裝 cyclegan 和 deformation net 的設計效用.
% 因為實際轉出來的並不是sem的圖,
% 所以不能在文字裡面宣稱我們可以把layout轉成sem
% 退而求其次用保守的說法寫: 
%                   系統大概念是把圖轉化去某個可作為媒介的domain

%先提示整個工作的主要目的.
%After the integrated circuit (IC) is fabricated on the (semiconductor) wafer according to the layout-generated GDS format file, generally speaking, SEM photography is used to check the quality of the IC.
%上面那句話放去intro
%或是用以下的格式放去intro
%Generally speaking, wafer foundries use SEM photography to check the quality of the integrated circuit (IC) after the IC is fabricated on the semiconductor wafer according to the layout. 

%\begin{figure*}
%%\begin{tabular}{ p{180pt}}
%\centering
%\includegraphics[width=0.95\textwidth,keepaspectratio=true]{figures/defnet.JPG} \par \\ [-0.8cm]
%\\
%%\end{tabular}
%\caption{The diagram of two-step framework.}
%\label{fig:fig02}
%\end{figure*}

%\begin{figure}
    %\centering
%    \includegraphics[width=0.48\textwidth,keepaspectratio=true]{figures/LayoutSEMHetero.JPG} %\par \\ [-0.2cm]
%    \caption{Two heterogeneous domains of images. (a) Layout designs. (b) SEM images.}
%    \label{fig:HeteroDomain}
%\end{figure}

As shown in Fig. \ref{fig:LithoNet}, LithoNet consists of a CycleGAN-based \cite{zhu2017unpaired} domain transfer network and a deformation prediction network. LithoNet is designed to learn how an IC fabrication process deforms the shape contours of a layout pattern. It can simulate the fabrication process to predict the  shape deformation for further virtual metrology applications based on i) a given layout and ii) a set of fabrication parameters.
%
%The proposed LithoNet is a two-step framework, as illustrated in Figure \ref{fig:DefNet} and aims to learn how an IC wafer fabrication process deforms the shape contours of a layout pattern so that it can accurately simulate the process to generate a synthetic SEM-styled image for further virtual metrology applications based on i) a given layout and ii) a set of fabrication parameters.
%%SEM-styled (scanning electron microscope) image from an input layout without actually going through the fabrication process ??
%%In order to score a risk assessment result of a layout design,%
One major difficulty in learning the shape deformation model between a layout pattern and its corresponding SEM image of fabricated circuitry lies in the fact that they are from heterogeneous domains. Specifically, an SEM image is a high-resolution, gray-scaled image with deep DOF (depth of field), whereas a layout is no more than a man-made binary pattern with only rectangular regional objects on it. As a result, the goal of LithoNet is to  predict the contour shapes by learning the pixel-wise shape correspondence between every paired layout and SEM images. Nevertheless, due to the poor contrast and scanning pattern noise in SEM images, it is usually  difficult to capture edge contours correctly from SEM images, on which a 1-pixel-drift corresponds to a nanometer-scale displacement on real IC products. Therefore, transferring the domain of SEM images to another intermediate domain without the above-mentioned contrast and noise problems would be beneficial.
 %cross-domain shape correspondence problem for layout and SEM images.

%\textbf{
%Therefore, the LithoNet needs to be trained via a two-step framework so that the difficulty in finding shape correspondence between two heterogeneous domains of images, i.e., layout and SEM image, can be overcome.
%}

%Our design concept of this two-phase framework is to convert SEM and layout images into an intermediate reference domain, where image contents are shaped into SEM-fashion with a layout-styled clear background.

 % of fabricated circuitry. 
To this end, we propose a two-step framework. In the first step, we use CycleGAN \cite{zhu2017unpaired} to transfer a gray-scale SEM image to an intermediate domain, where images have SEM-styled shape contours and layout-styled clear background.   
Then, in the second step, given a source layout along with fabrication parameters, LithoNet predicts the shape deformation introduced by the fabrication process. 
In sum, Step-I learns to bridge the gap between the SEM image and its binary layout  so that Step-II can learn the shape correspondence between the SEM image and its original layout.
In the following subsections, we will introduce our design in detail.

\subsection{Step I: Image domain transfer}
\label{subsec:301}

Because the SEM and layout images are of heterogeneous domains, we adopt an image domain transfer technique to \textit{align} their domains.
%We model the problem of SEM and layout images being of different domains as an image domain (style) transfer issue, and hence we adopt a style transfer technique to align the \textit{style} of SEM and layout images. 
By removing the interference introduced by the SEM imaging process, e.g., bias in brightness/contrast and scan-line noise, via CycleGAN \cite{zhu2017unpaired}, the processed SEM image is translated to the domain of the layout. That is, the processed SEM image retains its curvilinear shape boundaries yet is binarized as if it were a layout.

%描述training
%要在哪裡加入不能直接train style-converter、把layout變成sem的概念?
To this end, we train CycleGAN using i) a set of product-ICs' SEM images and ii) their associated segmentation masks.
The second set of images can be derived by applying either manual labelling,  advanced thresholding  techniques~\cite{saha2001optimum,otsu1979threshold}, interactive segmentation \cite{maninis2018deep,wang2018deepigeos}, or pseudo-background subtraction \cite{barnich2010vibe} on the source SEM images. 
Note that in order to guarantee the performance of domain transfer, segmentation masks with incorrect segmentation results are discarded under user-supervision.
Finally, we utilize the well-trained CycleGAN to transfer source SEM images into the layout style, and these processed SEM images are further taken as reference ground-truths to train LithoNet in Step-II.

Employing CycleGAN for domain transfer has two advantages. First, CycleGAN is an unpaired image-to-image translation method, and hence it can learn the majority decision of multiple image segmentation algorithms, including the analysis software provided by the SEM vendor, for SEM images based on a collection of segmentation results of different methods. 
Second, utilizing a "U-net Generator" to translate images, CycleGAN is essentially a U-net-based segmentation method~\cite{unet2015} supervised by its built-in "Discriminator" through an adversarial loss, thereby suggesting a more reliable segmentation result than U-net, a state-of-the-art segmentation benchmark. Additionally, we can simply discard some rare incorrect CycleGAN segmentation results by quick human-inspection to prevent LithoNet from learning incorrect shape correspondences.

%Therefore, the advantages of Phase-A are twofold. First, through style-transfer, the interference from the SEM imaging procedure, such as color-drift and scan-line noise, can be removed. Second, by transferring a gray-level SEM image into binary layout pattern in advance, Phase-2 only need to settle down a problem of transferring a binary image into another rather than the original complicated problem.

%We formulate the problem of removing differences between SEM image and the layout as an image style transfer issue. By removing the interference from the SEM imaging procedure, such as color-drift and scan-line noise in the background, a pre-processed SEM image can be regarded as of the same domain as its original layout. Therefore, we take CycleGAN as the preprocessing module in Phase-1.

%There are a lot of noise in original SEM images that make us can’t utilize these images directly. Thus we formulate it as an image style transfer problem and try to transfer the SEM image into the same domain as GDS layout. Here we use the cyclegan architecture shown in figure*. We train the cyclegan by collecting the SEM images as domain A and applying Otsu method on these SEM images to get binarized SEM images as domain B. Because not every binarized SEM images is good. So we pick the better binarized SEM images to train cyclegan. After we have preprocessed SEM images, we use these images as ground truth to train the deformation net.

\subsection{Step II: Shape Deformation Prediction}

To learn the shape correspondence and the deformation field between SEM and layout images, LithoNet is trained on a collection of image pairs, each containing a layout and a ground-truth segmentation mask, i.e., a processed SEM image, generated by Step-I described in Section \ref{subsec:301}. 
%\textbf{(write one more sentence to close this paragraph?)}

As shown in Fig. \ref{fig:LithoNet}, LithoNet consists of a generator and a warping module. 
The generator is a U-net~\cite{unet2015} like network that outputs a 2D dense correspondence map depicting the deformation field between the training image pairs. Then, using the sampling strategy used in the spatial transformer network (STN) \cite{jaderberg2015spatial}, the warping module synthesizes a warped version of the given input layout to simulate wafer-fabricated  circuitry based on the deformation map. 
%接下來要說這個設計的好處是甚麼 
STN is a differentiable module designed for enabling neural networks to actively spatially transform feature maps so that neural network models can learn invariance to translation, scale, rotation, and warping. Consequently, we adopt the sampling strategy of STN to benefit our LithoNet.

%===以下來自於STN paper abstract的原始敘述
%
%This differentiable module can be inserted into existing convolutional architectures, giving neural networks the ability to actively spatially transform feature maps, conditional on the feature map itself, without any extra training supervision or modification to the optimisation process. We show that the use of spatial transformers results in models which learn invariance to translation, scale, rotation and more generic warping, resulting in state-of-the-art performance on several benchmarks, and for a number of classes of transformations.

%R2Q2
Moreover, the deformation map $\mathcal{M}:\mathbb{R}^2\rightarrow\mathbb{R}^2$ describes the pixel-to-pixel displacement from a source layout image $\mathcal{S}$ to an SEM-styled image $\mathcal{J}$. Therefore, after LithoNet learns to predict the pixel-to-pixel correspondence, we apply the deformation map $\mathcal{M}$ on the layout $\mathcal{S}$ to derive the deformed shape contour. The warping process that relates $\mathcal{S}$, $\mathcal{J}$, and $\mathcal{M}$ can be expressed as 
$\mathcal{J}(m, n) = \mathcal{S}\big( \mathcal{M}^{-1} (m,n) \big)$, 
where $(m,n)$ denotes the pixel coordinate.

%Shown in Figure 6 is the magnitude of our deformation map. That is, for a given deformation map, we show its magnitude  as arrows pointing to the direction.

%寫舊作法不適合拿來檢驗製程中的global和local變化,
%兩個優點, 包裝成: 能夠明確輸出deformation field使得訓練結果能被visualize出來, 第二個透過這樣deformation field我們能容易觀察出global和local變形場之間的關係。
%至於我們會怎麼樣來控製變形場,則會在下一節描述。
In contrast to common image generation networks like \cite{isola2017image,wang2018discriminative}, the advantages of  LithoNet are twofold. 
First, LithoNet can generate and visualize a predicted deformation field, and therefore what has been learned by the network, i.e., the shape correspondences between input training image pairs, can be verified straightforwardly. 
Second, based on the visualized deformation field, it would be easier to identify possible impacts (e.g., defects), whether global or local, caused by the layout and the configuration parameters during fabrication process, on the physical appearance of an IC's metal layer.
%寫一句也能有助於我們了解製成參數、layout對IC成品的metal layer的外觀所造成的影響(無論全域或局部)
Concisely, the deformation field generated by our LithoNet is beneficial for clarifying both global and local shape correspondences between a layout and the SEM image of its product IC. 

%As for the cost function used to control the deformation net, we will describe it in next subsection.

\subsection{Training Loss Functions}

The training loss function $\mathcal{L}_{total}$ of LithoNet is primarily defined in the following form 
%We define a total loss $\mathcal{L}_{total}$ as a combination of four main terms,
\begin{equation}
%\mathcal{L}_{total} = \alpha \mathcal{L}_{rec} + \beta \mathcal{L}_{var} + \gamma \mathcal{L}_{smooth} + \eta \mathcal{L}_{reg} + \theta \mathcal{L}_{par}\mbox{.}
\mathcal{L}_{total} = \mathcal{L}_{rec} + \mathcal{L}_{var} + \mathcal{L}_{smooth} + \mathcal{L}_{reg} +  \mathcal{L}_{par}\mbox{.}
%Ltotal = αrecLrec+αvarLvar+αdsLds+αdrLdr, (1)
\label{eq:eq01}
\end{equation}
where, $\mathcal{L}_{rec}$ denotes the reconstruction loss that measures the dissimilarity between the training ground-truth $\mathcal{I}$ and the synthetic SEM-styled image $\mathcal{J}$. Meanwhile, $\mathcal{L}_{var}$ measures the variability difference between a paired training image pair, and $\mathcal{L}_{smooth}$ guarantees the smoothness of the deformation map. Finally, $\mathcal{L}_{reg}$ is used to penalize large displacements on the deformation map, and $\mathcal{L}_{par}$ is the regression loss of fabrication parameters. 

%Where Lrec encourages the reconstructed image to appear similar to the corresponding training ground truth, Lvar let the edge of layout in reconstructed image more conform with ground truth, Lds enforce smooth deformation map, Ldr prefer the deformation map’s value to be small. 

\noindent \textbf{A) Reconstruction Loss:}

The reconstruction loss term  $\mathcal{L}_{rec}(\mathcal{I}, \mathcal{J})$ is defined as the $L_1$ loss between the training ground-truth $\mathcal{I}$ and the synthetic SEM-styled image $\mathcal{J}$ as follows: % and the SSIM value. 
\begin{equation}
%\mathcal{L}_{rec}(\mathcal{I}, \mathcal{J}) = w_1 \parallel \mathcal{I} - \mathcal{J} \parallel_1 + w_2 \mbox{SSIM}(\mathcal{I}, \mathcal{J}) \mbox{.}
\mathcal{L}_{rec}(\mathcal{I}, \mathcal{J}) = \frac{1}{n}\parallel \mathcal{I} - \mathcal{J} \parallel_1  \mbox{,}
\label{eq:eq02}
\end{equation}
where $n$ denotes the number of pixels.
%Similar to the strategy used in the spatial transformer network (STN) \cite{jaderberg2015spatial}, we 
We derive $\mathcal{L}_{rec}$ by the following steps: 
%i) sampling densely pixel positions on the to-be-generated $\mathcal{J}$; ii) locating the correspondences of them on the input layout according to the deformation map; iii) using backward interpolation to estimate the sampled pixel values on $\mathcal{J}$; and finally, iv) generating an estimated $\hat{\mathcal{J}}$ via bilinear interpolation to calculate $\mathcal{L}_{rec}$. 
%上面的Line-149是增加bilinear, back-interp之前的原始敘述
i) densely sampling pixel positions on the to-be-generated $\mathcal{J}$; ii) locating the correspondences of them on the input layout according to the deformation map $\mathcal{M}$ that records the mapping relationship between pixels on $\mathcal{I}$ onto their counterparts on $\mathcal{J}$; iii) using backward interpolation to estimate the sampled pixel values on $\mathcal{J}$, i.e., $\hat{\mathcal{J}}(x,y) = \mathcal{I}(\hat{x}, \hat{y})$ with non-integer positions $(\hat{x}, \hat{y}) = \mathcal{M}^{-1}(x,y)$; and finally, iv) generating an estimated $\hat{\mathcal{J}}$ via bilinear interpolation\footnote{%
%====begin of footnote====
$\hat{\mathcal{J}}(x,y) = \mathcal{I}(\hat{x}, \hat{y})$  \\
$\approx 
\left[ \begin{array}{c}
    \left\lceil{\hat{x}}\right\rceil -\hat{x} \\ \hat{x}-\left\lfloor{\hat{x}}\right\rfloor
\end{array} \right]^t
\left[ \begin{array}{cc}
   \mathcal{I}(\left\lfloor{\hat{x}}\right\rfloor, \left\lfloor{\hat{y}}\right\rfloor) & \mathcal{I}(\left\lfloor{\hat{x}}\right\rfloor, \left\lceil{\hat{y}}\right\rceil)\\
   \mathcal{I}(\left\lceil{\hat{x}}\right\rceil, \left\lfloor{\hat{y}}\right\rfloor) & \mathcal{I}(\left\lceil{\hat{x}}\right\rceil, \left\lceil{\hat{y}}\right\rceil)
\end{array}
\right]
\left[ \begin{array}{c}
    \left\lceil{\hat{y}}\right\rceil -\hat{y} \\ \hat{y}-\left\lfloor{\hat{y}}\right\rfloor
\end{array} \right] \mbox{,}
$
where $\left\lceil \cdot \right\rceil$ and $\left\lfloor \cdot \right\rfloor$ denote \textit{ceiling} and \textit{floor} functions, respectively.
}
%====end of footnote====
to calculate $\mathcal{L}_{rec}$.

%During training, the network learns to generate an image by sampling pixel from input GDS layout image. Our image formation model references the image sampler from spatial transformer network (STN)* to sample the input image using $x$ and $y$ directions deformation map. We use bilinear sampling where the output pixel is weighted sum of two input pixels. Firstly, model sample the input GDS layout image according to the first channel of generator output which represents x direction’s deformation. Then repeat the same sampling step but using the second channel of generator output which represents y direction’s deformation.
%We use a combination of an $L_1$ and single scale SSIM term as our SEM image reconstruction loss Lrec,
%train的時候，網路會學習去產生一張影像、藉由從gds layout的sample點來學。我們的影像模型參考了STN的image sampler方式來對輸入影像進行x- y-方向的取樣。我們利用bilinear取樣. 第一,模型根據地一個channel鎖內涵的 x方向的變形場, 對layout影像進行取樣; 再來,利用同樣的方式對y方向作業, 我們利用L1norm和SSIM來代表我們的這一項的loss function.

\noindent \textbf{B) Total Variation Loss:}

The total variation loss $\mathcal{L}_{var}(\mathcal{I}, \mathcal{J})$ is defined as the total variation \cite{rudin1992nonlinear} of the \textit{signed} difference between $\mathcal{I}$ and $\mathcal{J}$, that is \\
\begin{equation}
%\mathcal{L}_{var}(\mathcal{I}, \mathcal{J}) = \mbox{TV}( \mathcal{I} - \mathcal{J} ) \mbox{.}
\mathcal{L}_{var}(\mathcal{I}, \mathcal{J}) = \sum |\nabla ( \mathcal{I} - \mathcal{J} )| \mbox{.}
\label{eq:eq03}
\end{equation}

This term is designed to align the shape contours of $\mathcal{J}$ with those of $\mathcal{I}$. Without this term, the loss function might be dominated by the reconstruction loss described in (\ref{eq:eq02}), and consequently LithoNet would generate a bizarre synthetic image $\mathcal{J}$, which can produce a high overlap ratio compared with ground-truth image $\mathcal{I}$ but has unnaturally jiggling contours. 
In other words, $\mathcal{L}_{var}$ aims to retain the shape similarity.
%https://blog.csdn.net/weixin_42447651/article/details/82990941
%https://blog.csdn.net/afgh2587849/article/details/6401181

%In our observations, only using reconstruction loss often leads to badly simulation results at edge of layout. We think that is because reconstruction loss only considers the whole image loss. The edge information will been average by the region inside layout or background. Thus we calculate the total variation loss Lvar by computing total variation of difference between generated image and ground truth.

\noindent \textbf{C) Smoothness Loss} 
%The smoothness loss is again a penalty term. Hence, we define the smoothness loss as the $L_1$-norm of the gradient of deformation map. That is, 
%\begin{equation}
%\mathcal{L}_{smooth} = \parallel \nabla \mathcal{M} \parallel_1 \mbox{.} %\mbox{,}
%\label{eq:eq04}
%\end{equation}
%%where $\mathcal{M}$ is the deformation map, and $\nabla$ denotes gradient operator.
%Note that because contour edges on the layout always result in discontinuities on the deformation map $\mathcal{M}$, this smoothness loss should multiplies by an edge-aware weighting factor, i.e., the $\gamma$ in Equation (\ref{eq:eq01}), to suppress the unnecessary penalty contributed by contour edges. In addition, such edge-aware weighting factor can be derived according to the gradient of both layout and SEM images.
%

The smoothness loss is a penalty term defined as the $L_1$-norm of the weighted gradient of the deformation map:
\begin{equation}
\mathcal{L}_{smooth} = \parallel (\nabla \mathcal{M}) \circ \mathbf{W} \parallel_1 \mbox{,}
\label{eq:eq04}
\end{equation}
where $\circ$ denotes the Hadamard product, % (entry-wise matrix multiplication), 
and $\mathbf{W}$ is an edge-aware weighting matrix defined as
\begin{equation}
\mathbf{W}(x,y) = e^{-\left( | \nabla \mathcal{S}(x,y) | + | \nabla \mathcal{I}(x,y) | \right) } \mbox{.}
\label{eq:eq04b}
\end{equation}

Note that contour edges on the input layout $\mathcal{S}$ and the ground-truth layout-styled SEM image $\mathcal{I}$ result in discontinuities in the deformation map $\mathcal{M}$. Because such discontinuities contribute to an unnecessary smoothness penalty, $\mathcal{L}_{smooth}$ should be suppressed appropriately according to the gradient information of both layout and SEM images. 

%We encourage deformation map to be smooth at adjacent pixels with a $L_1$ penalty on the gradient of deformation map. Meanwhile, because deformation discontinue often occur at layout edges, so we weight this loss with an edge-aware term using the gradient of GDS layout image and SEM image.

\noindent \textbf{D) Regularization Loss} 

The regularization loss is defined as the $L_1$ norm of deformation map $\mathcal{M}$:
\begin{equation}
\mathcal{L}_{reg} = \parallel \mathcal{M} \parallel_1 \mbox{.}
\label{eq:eq05}
\end{equation}

This term reflects the fact that the deformation caused by wafer fabrication tends to be small, as will be discussed in Section \ref{subsec:406}.
%as will be shown in Figure \ref{fig:fig07}.
%In general, the deformation between GDS layout and SEM image is small. That is mean the pixel values of deformation map must be small too. Therefore, we calculate a $L_1$ norm of deformation as a deformation regularization loss.

\noindent \textbf{E) Regression Loss for Fabrication Parameters} 

Because the configuration parameters of a fabrication process are continuous variables that influence the physical appearance of the wafer layer, we formulate the relationship between the fabrication parameters and the appearance of wafer layer as a regression problem. The regression loss $\mathcal{L}_{par}$ is defined as 
\iffalse
\begin{equation}
\mathcal{L}_{par} = 
\underbrace{\parallel D_y( G(\mathcal{S}|y_0 ) ) - y_0 \parallel_2 }_{\mbox{Generator loss}}
+\underbrace{\parallel D_y( \mathcal{I}_{y_i} ) - y_i \parallel_2}_{\mbox{Discriminator loss}}   \mbox{.}
\label{eq:eq06}
\end{equation}
\fi
%\textbf{Therefore, how do we implement and define this term?}
\begin{equation}
\mathcal{L}_{par} = 
\underbrace{\parallel D_y( G(\mathcal{S}|y ) ) - y \parallel_2 }_{\mbox{Generator loss}}
+\underbrace{\parallel D_y( \mathcal{I}_{y} ) - y \parallel_2}_{\mbox{Discriminator loss}}   \mbox{.}
\label{eq:eq06}
\end{equation}
where $\mathcal{I}$ is the ground-truth shape segmented from the SEM image used for training; $y$ is the fabrication parameter vector corresponding to $\mathcal{I}_{y}$; %$\mathcal{S}$ and $y_0$ respectively denote the input layout and input fabrication parameter vector for prediction; and, $G(S|y)$ is the predicted deformed shape. 
$\mathcal{S}$ denotes the input layout; and, $G(S|y)$ is the predicted deformed shape. 
Therefore, this loss term aims to train i) a generator able to predict a synthesized SEM-styled image based on the given $\mathcal{S}$ and $y$, and %ii) a discriminator able to estimate the fabrication parameter vector $y$ associated with $\mathcal{I}_{y}$. 
ii) a discriminator able to discriminate whether each entry of the extracted parameter vector $D_y(\mathcal{I}_y)$ is identical to the corresponding entry within the groundtruth fabrication parameter vector $y$. 
%Note that because the generator loss and the discriminator loss are calculated independently during the training process, we use two different symbols, i.e., $y_0$ and $y_i$, in (\ref{eq:eq06}) to specify this situation; however, $y_0$ equals to $y_i$ for every training sample pair in the LithoNet design.

% 
% S應該說是輸入的layout, G(S| y0)是根據輸入layout + 參數所合成出來的SEM圖
% I則是訓練generator時所用的Layout ground-truth, y_i是已知的參數.
% 上文還沒加入對變數的解釋,也還沒確認y_i, y_0的詳細寫法,要小心.
%
%Because the process parameter is continuous, so we use a regression model that try to regress different SEM image and train the model by Process Parameter Regression Loss L

%We named the whole network structure in Phase-2 \textit{deformation-net}.
%
%
%
%
%remove interference from the SEM imaging processing
%
%
%
%the phase-1 was design to make style-transfer so that  first to remove 
%
%For the former phase, its goal is to transform an input GDS layout image 
%
%
%to remove interference from the SEM imaging processing

%------------------------------------------------------------------------	
	\section{OPCNet: A CNN-Based Photomask Corrector}
	\label{sec:opcnet}
	%\subsection{Architecture of Photomask Generator}
%\label{subsec:photomask}

As described in Section \ref{subsec:previousMask}, the major challenge in developing a learning-based mask optimizer is to collect a comprehensive amount of ground-truth mask data, e.g. well OPC-corrected photomasks of various layout patterns, leading to desired shapes of fabricated circuitry. 
%==下面一句是林老師的原版備份
%As described in Section \ref{subsec:previousMask}, the major challenge in developing a learning-based mask optimizer is to collect a comprehensive amount of ground-truth mask data corresponding to various layout patterns, e.g., well OPC-corrected photomasks leading to desired shapes of fabricated circuitry. 
This is, however, very costly and time-consuming. To overcome this difficulty, as shown in Fig. \ref{fig:framework}(b), we utilize a pre-trained LithoNet as an auxiliary module to train our photomask optimizer, i.e., OPCNet. Given an IC layout pattern, OPCNet aims to predict an OPC-corrected mask pattern so that, after being deformed by the lithography and etching processes that are simulated by LithoNet, the predicted deformed shape will be as close as possible the original layout pattern. 
%
% ==== 舊版刪除 ====
%Therefore, OPCNet can be regarded as the  inverse model of LithoNet. As a result, for a desired layout pattern, we can use its predicted outputs of LithoNet as the input of OPCNet, and the desired layout itself as the corresponding output of OPCNet. 
% ==== 以上舊版刪除,以下為新版 ====
By regarding the LithoNet-OPCNet network as a composite function $f=h \circ g$ with $h$ and $g$ denoting respectively LithoNet and OPCNet, this design can be expressed as $\min \| \mathcal{S} - f(\mathcal{S}) \|$, where $\mathcal{S}$ and $f(\mathcal{S})$ are respectively the input layout and the final prediction produced by the LithoNet-OPCNet network.
Therefore, because such minimization optimizes to $f=1$, which implies $h \circ g = 1$, OPCNet and LithoNet should be inverse functions of each other. As a result, for a desired layout pattern, we can use the predicted output of OPCNet as the input of LithoNet, and the desired layout itself as the corresponding input of OPCNet. Consequently, we can train OPCNet without the need for collecting the ``ground-truth" OPC-corrected photomasks.
%
%
%Given a collection of such input-output pairs, 

Specifically, given a layout design pattern $\mathcal{S}$, OPCNet aims to generate a photomask $\mathcal{K}$, whose lithography and etching simulation result $\mathcal{J}$ predicted by LithoNet best matches $\mathcal{S}$. This design makes our OPCNet ``groundtruth-free'' during the training stage, assuming LithoNet is already well-trained. In addition, with the design of the \textit{input-output consistency loss} used to measure the dissimilarity between a layout design pattern $\mathcal{S}$ and its lithography simulation result $\mathcal{J}$, OPCNet becomes a self-supervised learning method. The whole pipeline of our mask optimization method is illustrated in Fig. \ref{fig:framework}(b). Note that i) the pretrained LithoNet is fixed while training OPCNet, and ii) OPCNet is intrinsically a generator for translating a layout pattern $\mathcal{S}$ into its optimal photomask $\mathcal{K}$ based on the wafer fabrication model learned by LithoNet.
%After LithoNet training is complete and converge, we exploit an additional generator followed by the LithoNet to learn mask optimization so that the generator can translate a layout image into optimal mask that its wafer image is expected as similar as input layout. The whole pipeline is shown in Figure \ref{fig:maskpipline}.

% \begin{figure}
    %\centering
%    \includegraphics[width=0.48\textwidth,keepaspectratio=true]{figures/pipelineOPCnet.JPG}
%    \caption{Pipeline for mask optimization.}
%    \label{fig:maskpipline}
%\end{figure}

%Recently, some deep learning based methods like [14, 16], they need to use ground truth optimal mask data to train their network. However, we proposed a novel training objective function called input-output consistency loss to achieve unsupervised learning. That is, we only utilize the layout images to train the mask generator. Note that when training the mask generator, the pre-trained LithoNet is fixed and only doing backpropagation to update the mask generator.

%=======================================
\subsection{Application Scenarios}

The LithoNet-OPCNet network can serve two purposes. First, when LithoNet is well pre-trained on a comprehensive set of (layout, SEM) pairs if during IC fabrication no OPC is performed, or on a set of (mask, SEM) pairs if OPC is performed, LithoNet can accurately predict the shape deformations due to the lithography and etch processes. Since OPCNet is the inverse function of LithoNet, it can be used to predict the OPC-optimized mask for a target layout pattern that would minimize the discrepancy between the fabricated IC shape and the target layout pattern without the need for collecting ''ground-truth" OPC-optimized masks. In this way, the LithoNet-OPCNet network potentially can replace the function of current OPC prediction models.

Second, if the training samples are not comprehensive enough to train a fully reliable LithoNet model, the LithoNet-OPCNet may not be able to completely  replace current OPC prediction models. However, if LithoNet can achieve a reasonable accuracy, the LithoNet-OPCNet network can still be used to verify if there is any inconsistency between the optimized mask prediction and the conventional OPC mask---an obvious inconsistency implies the fab-plant need to update the OPC model by collecting specific process-window conditions over the input layout structure.

\subsection{Training Loss Functions for OPCNet}
\label{subsec:OPCloss}

The overall training loss $\mathcal{L}_{\mathcal{K}}$ of OPCNet is defined as
\begin{equation}
    \mathcal{L}_{\mathcal{K}} = \mathcal{L}_{IO} + \mathcal{L}_{Kvar} + \mathcal{L}_{Ksmooth} \mbox{,}
    \label{eq:maskloss}
\end{equation}
where $\mathcal{L}_{IO}$ denotes the input-output consistency loss measuring the dissimilarity between input layout $\mathcal{S}$ and LithoNet's output  $\mathcal{J}$, 
$\mathcal{L}_{Kvar}$ represents the total variation loss on the difference between $\mathcal{S}$ and $\mathcal{J}$, and
 $\mathcal{L}_{Ksmooth}$ denotes the mask smoothness loss for ensuring the smoothness of the obtained photomask patterns $\mathcal{K}$.

\noindent \textbf{A) Input-Output Consistency Loss:}

The input-output consistency loss $\mathcal{L}_{IO}(\mathcal{S}, \mathcal{J})$ aims to guide the learning of OPCNet so that the shape predicted by LithoNet $\mathcal{J}$ best matches the desired input layout $\mathcal{S}$, provided that the source layout is OPC-corrected by the learned OPCNet. The loss term is defined as follows: 
\begin{equation}
    \mathcal{L}_{IO}(\mathcal{S}, \mathcal{J}) = \frac{1}{n} \| \mathcal{S} - \mathcal{J} \|_1 \mbox{,}
    \label{eq:Kconsist}
\end{equation}
where $n$ denotes the number of pixels. %With this loss term, the mask generator can learn how to revise the input layout to generate an optimized mask by unsupervised learning.

\noindent \textbf{B) Total Variation Loss:}

Similar to (\ref{eq:eq03}), the total variation loss $\mathcal{L}_{Kvar}(\mathcal{S}, \mathcal{J})$ is defined as the total variation of signed difference between the input layout $\mathcal{S}$ and the prediction of LithoNet $\mathcal{J}$:
\begin{equation}
    \mathcal{L}_{Kvar}(\mathcal{S}, \mathcal{J}) = \sum |\nabla(\mathcal{S}-\mathcal{J}) | \mbox{,}
    \label{eq:Kvar}
\end{equation}
which is again an empirical term used to avoid unnatural patterns on the predicted shapes. 
%This term is primary consider the shape contours of $\mathcal{S}$ and $\mathcal{J}$. 
%$\mathcal{L}_{Kvar}$ plays a role similar to $\mathcal{L}_{Kvar}$ described in (\ref{eq:eq03})
$\mathcal{L}_{Kvar}$ prevents $\mathcal{L}_\mathcal{K}$ from being dominated by the I/O-consistency loss $\mathcal{L}_{IO}$. Without this term, the OPCNet may produce a unnatural correction.
%Without this term, the total loss function $\mathcal{L}_{mask}$ would be dominated by the input-output consistency loss described in (\ref{eq:Kconsist}) and lead to a poor prediction. 

\noindent \textbf{C) Mask Smoothness Loss:}

The mask smoothness loss is defined to be the $L_1$-norm of the gradient of the mask prediction, that is,
\begin{equation}
    \mathcal{L}_{Ksmooth} = \| \nabla \mathcal{K} \|_1 \mbox{.}
    \label{eq:Ksmooth}
\end{equation}

This term penalizes the discontinuity on the corrected photomask $\mathcal{K}$ to guarantee the smoothness of shape contours of $\mathcal{K}$. 
%Compared with $\mathcal{L}_{smooth}$ in (\ref{eq:eq04}), $\mathcal{L}_{Ksmooth}$ does not collaborate with an edge-aware weighting matrix
Note that $\mathcal{L}_{Ksmooth}$ does not incorporate with an edge-aware weighting matrix, since there are no ground-truth masks that define true contour edges in the training dataset.

In practice, there are some restrictions on what kind mask shapes can be made by a mask shop.  We can integrate such mask manufacturing rules checking (MRC) with OPCNet in two ways: (1) formulating the MRC as training loss functions  of OPCNet, or (2) using a post-processing step based on the MRC rules to modify the OPC-corrected layout patterns generated by OPCNet. The second method is commonly used in practice, but OPCNet has the capability to adopt the first method or a combination of the two methods.

%------------------------------------------------------------------------	
	\section{Experimental Results}
	\label{sec:experiments}
	\begin{figure}[t]
    \centering
    \includegraphics[width=0.48\textwidth,keepaspectratio=true]{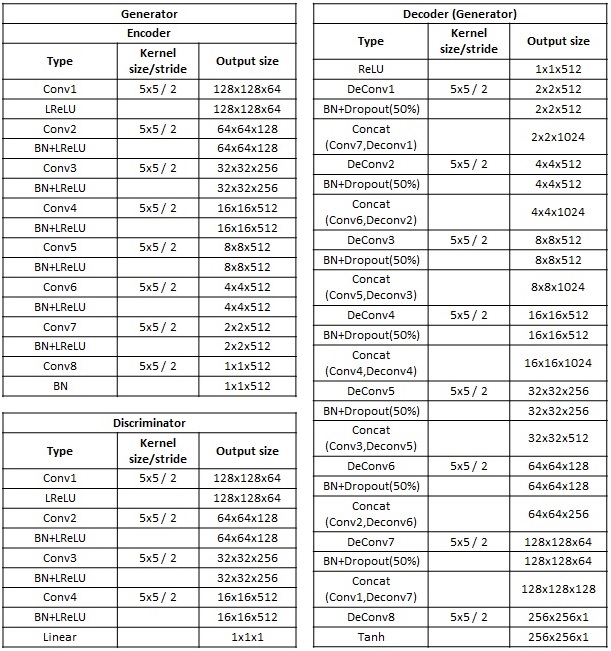}
    \caption{
    Network Architecture of LithoNet. Its generator consists of an encoder and a decoder. OPCNet is architecturally identical to LithoNet's generator. 
    }
    \label{fig:Architecture}
\end{figure}

\subsection{Dataset and Settings}
\label{subsec:401}

Images demonstrated in this work are selected from two datasets provided by United Microelectronics Corporation (UMC). 
Both these two UMC datasets consist of pairs of images, each containing one layout image patch and its wafer's SEM image patch. UMC dataset \#1 contains SEM images taken from wafers fabricated with the same fabrication parameters, and UMC dataset \#2 contains SEM images taken from wafers fabricated with \textit{seven} various normalized parameter settings ranging from $-0.9$ to $+0.9$. 
%The first dataset contains (i) a \textbf{942}-pair training subset, and (ii) a \textbf{100}-pair blind testing subset; whereas the second dataset consists of (i) a 7399
In total, UMC dataset \#1 contains (i) a 928-pair training subset and (ii) a 114-pair blind testing subset, whereas UMC dataset \#2 contains (i) a subset comprising $1057\times7$ pairs\footnote{there are 1,057 layouts and 7 different settings per layout, so 7,399 pairs of images in total.} for training and (ii) another subset comprising $12\times7$ pairs for blind testing. All images in the blind testing set are collected from historical fabrication data. Compared with those in the training sets, the blind test images are of much larger sizes and contain unseen design patterns. 
We trained CycleGAN for style-transfer in Step-I  on UMC dataset \#1, and LithoNet on UMC datasets \#1 and \#2. 
As for OPCNet, it was trained on paired data, each of which contains (i) a layout image $\mathcal{S}$ in the first dataset and (ii) its fabricated IC shape $\mathcal{J}$ predicted by feeding $\mathcal{S}$ into a pre-trained LithoNet. %As a result, OPCNet can be trained in a self-supervised manner. 
In our experiments, all image patches are downscaled from $512 \times 512$ to  $256 \times 256$ to reduce the computational complexity. 
Each $512 \times 512$ source image corresponds to a $2\times2 \mu m^2$ region, so aliasing will not occur in this case. 
The five loss terms described in (\ref{eq:eq01}) are weighted empirically by $(100, 0.001, 150, 0.002, 10)$. Meanwhile, the weighting coefficients for OPCNet described in (\ref{eq:maskloss}) are (50, 0.001, 50). 
These weighting coefficients are determined according to the following two steps. First, because the reconstruction loss and the smoothness loss in (\ref{eq:eq01}) and (\ref{eq:maskloss}) are more considerable than the others, we assign them with larger weighting coefficients and adjust the weighting coefficients until reaching reasonable results. In this step, the coefficients of other loss terms are temporarily set to be zero. Second, we assign the other loss terms with much smaller coefficients initially and then adjust them to make the training process easily converge.

\begin{figure}
    \centering
    \includegraphics[width=0.48\textwidth]{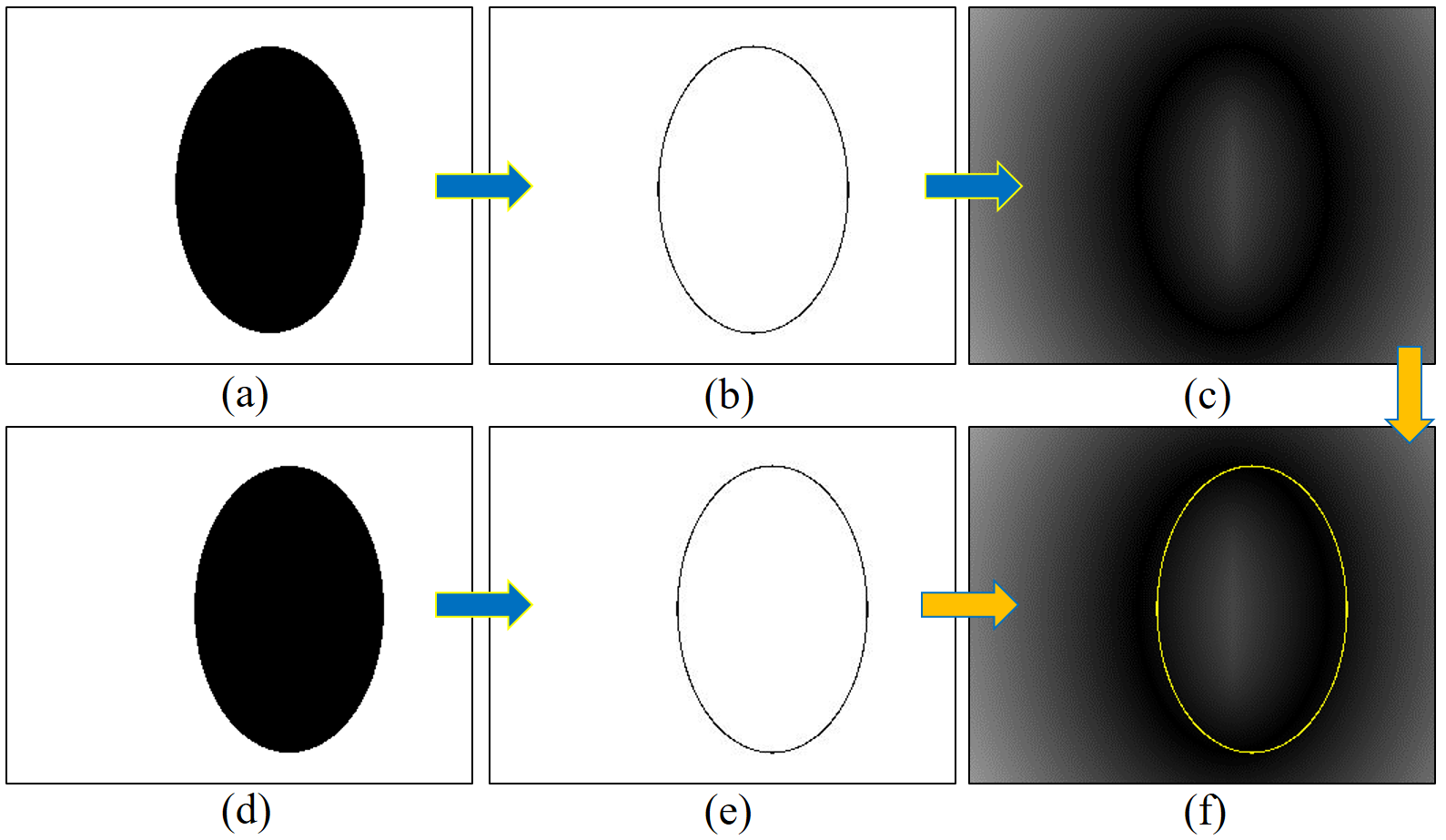}
    \caption{ Illustration of contour-to-contour distance (C2Cdist). (a) Ground-truth (GT); (b) the contour of GT; (c) the distance map \cite{maurer2003linear} of GT's contour obtained by MATLAB function \textit{bwdist}; (d) the input; (e) the contour of the input; (f) the overlay of (e) on GT's distance map. Then, \textit{C2Cdist} can be derived by averaging distance values collected along the input's contour.%Collecting distance values along input's contour to derive averaged Input-to-GT distance.
    }
    \label{fig:DistC2C}
\end{figure}

\begin{figure}
\begin{tabular}{ p{360pt}}
\includegraphics[width=0.48\textwidth,keepaspectratio=true]{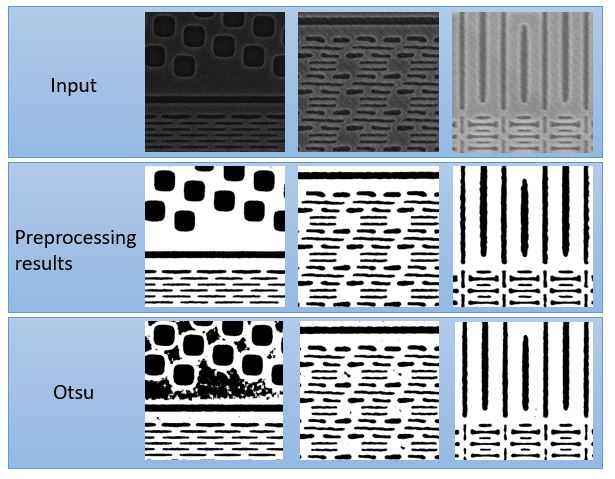} %\\ [-0.4cm]
\vspace{-0.8cm}
\end{tabular}
\caption{Comparison between the segmentation masks obtained by  CycleGAN \cite{zhu2017unpaired} trained on UMC dataset \#1 and traditional Otsu thresholding.}
\label{fig:segmentation}
\end{figure}

\begin{figure*}[!t]
%\begin{tabular}{ p{400pt}}
\centering
\includegraphics[width=0.97\textwidth,keepaspectratio=true]{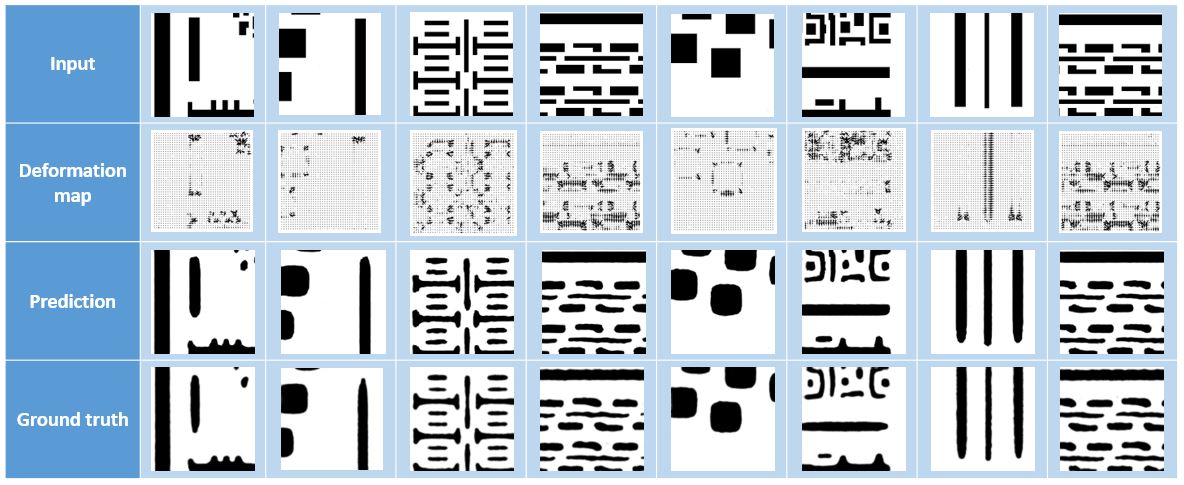} %\par \\ 
%\end{tabular}
\caption{Comparison of the input layout patterns, the predicted deformation maps, the predictions of fabricated IC shapes based on the deformation maps, and the ground-truths of fabricated IC shapes extracted from their associated SEM images. 
The second row illustrates every deformation map $\mathcal{M}(m,n) = x_m\hat{i}+y_n\hat{j}$ as its per-pixel magnitude $\sqrt{x_m^2+y_n^2}$ pointing to the deformation direction $\hat{v}=(x_m\hat{i}+y_n\hat{j})/\sqrt{x_m^2+y_n^2}$.}
%\caption{Predicted deformation maps and the comparison between prediction and ground truth.}
\label{fig:deformation_map}
\end{figure*}

%We use two datasets collected by a semiconductor Corporation. First dataset has total 1042 pairs image and each pair have a GDS layout image and an SEM image. This dataset is divided into two sets, 942 pairs for training and 100 pairs for testing. We utilize this dataset to train the image style transfer for image preprocessing and the deformation net without attribute. 
%Second dataset has total 7399 pairs image of seven different process parameter attributes. We keep * pairs for training and the rest for testing. We use this dataset to train the deformation net with attribute conditions.
%The size of the images in both datasets is $512 \times 512$ pixels. But in order to speed up the operation time, we resize the images from $512 \times 512$ to $256 \times 256$ pixels.

\subsection{Architecture and Run-time Information}
\label{subsec:arch}

Fig. \ref{fig:Architecture} shows the architectures of subnetworks constituting LithoNet, including i) the encoder of the generator, ii) the decoder of the generator, and iii) the discriminator. OPCNet shares the same architecture as LithoNet's generator. 
On average, LithoNet and OPCNet take 0.0156 and 0.0150 seconds to run a simulation on a $256\times256$ image on an NVIDIA 2080Ti GPU, respectively. The whole training process takes about 1.5 days on a server equipped with one NVIDIA P100 GPU. Note that on the server, it takes about 34 seconds to run OPC contour simulation for a $4\times1.7 \mu m^2$ layout patch.

\subsection{Performance Metrics}
\label{subsec:402}

The performance of our model is evaluated objectively in terms of  some widely-used similarity metrics, including Intersection Over Union (IOU), SSIM \cite{SSIM}, and per pixel error rate. 
%SSIM is a perception-based model that considers image degradation as perceived change in structural information, and currently it is one of the standard built-in metrics in environments like MATLAB (\textit{ssim}) and Python (\textit{skimage.metrics.structural\_similarity}). As for IOU and per pixel error rate (ErrorRate), they 
These three metrics are defined below.
\begin{eqnarray}
IOU(x, y) &=& \frac{\cap(x, y)}{\cup(x, y)} \mbox{,}\\
ErrorRate &=& \frac{FP+FN}{TP+TN+FP+FN} \mbox{, and}\\
SSIM(x, y) &=& \frac{(2\mu_x \mu_y + C1)(2\sigma_{xy}+C2)}{(\mu_x^2+\mu_y^2+C1)(\sigma_x^2 + \sigma_y^2 + C2)} \mbox{,}
\end{eqnarray}
where $\cap$ and $\cup$ denote respectively set intersection and set union; and, TP, TN, FP, and FN stand for true positive, true negative, false positive, and false negative, respectively. The SSIM index %, one of the built-in metrics in environments like MATLAB and Python,
measures the structural similarity between two images. In the equation above, $\mu_x$ and $\sigma_x$ denote the average and the variance of image $x$, $\sigma_{xy}$ denotes the covariance, and $C_1$ and $C_2$ are variables stabilizing the division.

Finally, we also utilize the contour-to-contour distance, hereafter abbreviated as C2Cdist, to approximate the \textit{Edge placement Error} (EPE) and the \textit{Edge Displacement Error} used in \cite{ye2019lithogan}. This metric, methodologically similar to EPE, measures the mean contour-to-contour distance between a lithography prediction and its SEM contour ground-truth. We utilize this strategy because an SEM prediction usually contains multiple irregular regions whose bounding boxes may be overlapped, and thus bounding boxes cannot suggest a fair distance measure for the whole SEM prediction. The \textit{C2Cdist} metric, measured in pixels, is illustrated in Fig. \ref{fig:DistC2C} and available for download at \cite{C2Cdist}. 
%
% \textbf{[cite something]}.
We will demonstrate in detail that our model outperforms other image-to-image translation methods and the standard OPC approach. % used in semiconductor virtual-metrology. 

%Evaluation of image to image translation task is a difficult problem. In our experiments we use Intersection Over Union (IOU), Structural Similarity (SSIM)*	 imagesand per pixel error rate to evaluate the performance of our model.  Besides, we also present the visual results that can be judge by human perceptual. \\
%*Image quality assessment: from error visibility to structural similarity

\subsection{LithoNet}
\label{subsec:Litho_main}

\subsubsection{Image domain transfer}
\label{subsec:403}

In Fig. \ref{fig:segmentation}, we compare our image domain transfer results with images derived by the traditional Otsu thresholding method \cite{otsu1979threshold}. Obviously, the source SEM images contain typical complications from the SEM imaging process, such as bias in brightness/contrast probably due to gain-shift and scanning-pattern noise. It is thus difficult for common  methods to threshold an SEM image appropriately. 
%By style-transferring an SEM image via CycleGAN \cite{zhu2017unpaired}, we can successfully obtain an binarized SEM image with its contour shape keeping unchanged.
By exploiting a well-trained translator, e.g., CycleGAN \cite{zhu2017unpaired}, an SEM image can be transferred into a layout-styled format with its contour shapes unchanged.

%In sum, this experiment demonstrates that 
%%through our design of an intermediate domain, \
%a noisy SEM image can be transferred into a binarized layout-styled format in a robust way. 

%In figure* shows the first step image style transfer results compared with traditional binarizaion method*. Obviously, the results show that the cyclegan is more robust to noise, color shift and scanning pattern and successfully translates the original SEM images to layout-styled SEM images.
\subsubsection{Prediction Results}
\label{subsec:406}

Fig. \ref{fig:deformation_map} illustrates the deformation map predicted from the input layout, the predictions of fabricated IC shapes based on the deformation map, and the corresponding ground-truths of fabricated IC shapes extracted from their associated SEM images.
The deformation maps show that LithoNet successfully learns to widen lines within open areas and to condense lines otherwise. Because such information is the key to the metrology applications, such as layout scoring and OPC simulation described in Fig. \ref{fig:metrology}, this experiment also demonstrates that LithoNet can be used to bridge computer vision techniques with both fields of semiconductor manufacturing and computer-aided-design.

\subsubsection{Ablation Study of Loss Terms}
\label{subsec:404}

\begin{figure}[t]
    \centering
    \includegraphics[width=0.48\textwidth,keepaspectratio=true]{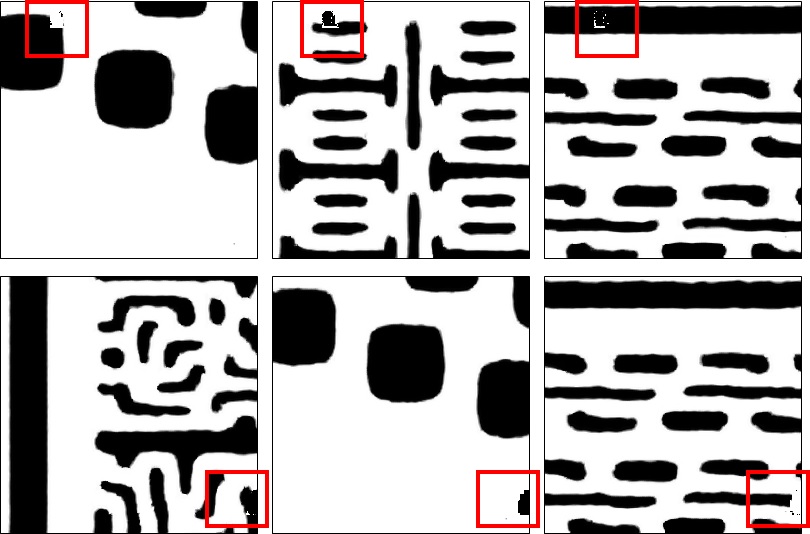}
    \caption{Prediction results by LithoNet trained on UMC dataset \#1 without the smoothness loss term $\mathcal{L}_{smooth}$.}
    \label{fig:withoutsmooth}
\end{figure}

Here we examine and discuss the effectiveness of individual loss terms in (\ref{eq:eq01}). 
First of all, we made numerical comparisons among different loss settings in Table \ref{table:MStable01} and Table \ref{table:MStable02}, each of which corresponds to a different dateset. The values in parentheses are final loss values on training set during training. The results shown in Table \ref{table:MStable01} were derived by LithoNet trained on UMC dataset \#1, %large dataset consisting of 942 training samples; 
whereas Table \ref{table:MStable02} shows the performance of LithoNet trained on a small subset of UMC dataset \#1 containing 480 training patches (obtained from 16 image samples by using only overlapped-cropping to obtain 30 patches for each sample for data augmentation). From Tables \ref{table:MStable01} and \ref{table:MStable02}, we can observe that the total-variation loss, $\mathcal{L}_{var}$, contributes significantly to the performance improvement. Moreover, $\mathcal{L}_{smooth}$ is beneficial to improve the objective performance when only a very limited amount of training samples is provided, as shown in Table \ref{table:MStable02}. On the contrary, as listed in Table \ref{table:MStable01}, $\mathcal{L}_{smooth}$ contributes less effectively to the objective performance when a comprehensive enough training dataset is given. We demonstrate the SEM-styled images predicted according to small training dataset without using the smoothness loss $\mathcal{L}_{smooth}$ in Fig. \ref{fig:withoutsmooth}, where unexpected artifacts are highlighted in red rectangles. This experiment set shows the necessity of $\mathcal{L}_{smooth}$, especially in cases of a small training set. 

\begin{table}[t]
\centering
\caption{\small Ablation study of different loss settings on UMC dataset \#1 (Data in parentheses are from training set.)}
\begin{tabular*}{\linewidth}{l}
\includegraphics[width=0.48\textwidth]{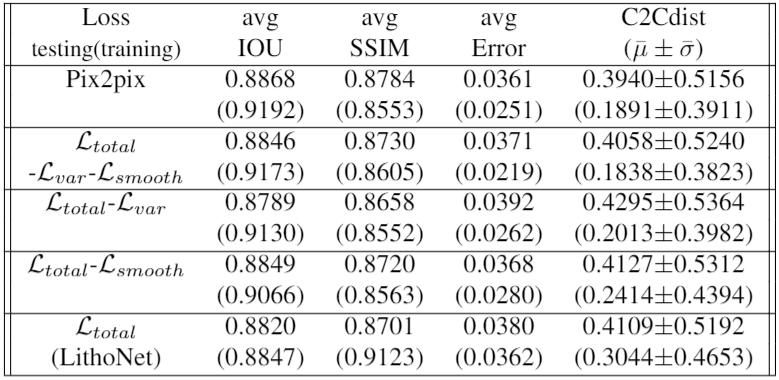}
\end{tabular*}
\label{table:MStable01}
\end{table}
\begin{table}[t]
\centering
\caption{\small Ablation study of different loss settings on a small subset of UMC dataset \#1 %(Parentheses denote training data.)
}
\begin{tabular}{l}
\includegraphics[width=0.48\textwidth]{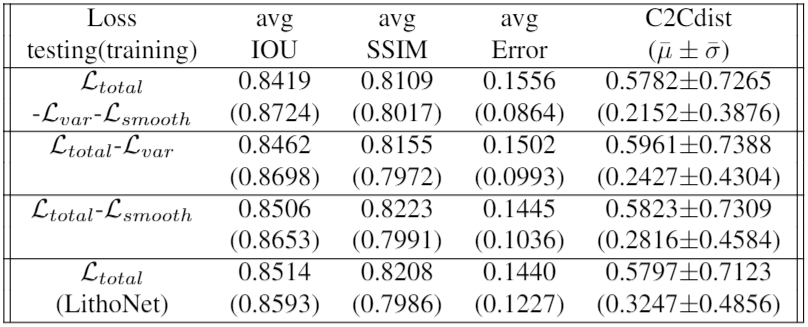}
%\begin{tabular}{cccccl}
%\cline{1-5}
%\multicolumn{1}{||c}{Loss} & avg & avg & avg &   \multicolumn{1}{c||}{C2Cdist} & \\ 
%\multicolumn{1}{||c}{testing(training)} & IOU & SSIM & Error &  \multicolumn{1}{c||}{($\bar{\mu}\pm\bar{\sigma}$)} & \\ \cline{1-5}
%\multicolumn{1}{||c}{\small{$\mathcal{L}_{total}$}}
%                         & 0.8419 & 0.8109 & 0.1556 &  \multicolumn{1}{c||}{0.5782$\pm$0.7265} & \\ %\cline{1-6}
%\multicolumn{1}{||c}{\small{-$\mathcal{L}_{var}$-$\mathcal{L}_{smooth}$}}
%                         & (0.8724) & (0.8017) & (0.0864) &  \multicolumn{1}{c||}{(0.2152$\pm$0.3876)} & \\ \cline{1-5}
%\multicolumn{1}{||c}{\small{$\mathcal{L}_{total}$-$\mathcal{L}_{var}$}}
%                         & 0.8462 & 0.8155 & 0.1502 & \multicolumn{1}{c||}{0.5961$\pm$0.7388} & \\ %\cline{1-6}
%\multicolumn{1}{||c}{}
%                         & (0.8698) & (0.7972) & (0.0993) &  \multicolumn{1}{c||}{(0.2427$\pm$0.4304)} & \\ \cline{1-5}
%\multicolumn{1}{||c}{\small{$\mathcal{L}_{total}$-$\mathcal{L}_{smooth}$}}
%                         & 0.8506 & 0.8223 & 0.1445 &  \multicolumn{1}{c||}{0.5823$\pm$0.7309} & \\ %\cline{1-6}
%\multicolumn{1}{||c}{}
%                         & (0.8653) & (0.7991) & (0.1036) &  \multicolumn{1}{c||}{(0.2816$\pm$0.4584)} & \\ \cline{1-5}
%\multicolumn{1}{||c}{\small{$\mathcal{L}_{total}$}}
%                         & 0.8514 & 0.8208 & 0.1440 &  \multicolumn{1}{c||}{0.5797$\pm$0.7123} & \\ %\cline{1-6}
%\multicolumn{1}{||c}{\small{(LithoNet)}}
%                         & (0.8593) & (0.7986) & (0.1227) &  \multicolumn{1}{c||}{(0.3247$\pm$0.4856)} & \\ \cline{1-5}
\end{tabular}
\label{table:MStable02}
\end{table}

\begin{figure}[t]
\centering
\includegraphics[width=0.48\textwidth,keepaspectratio=true]{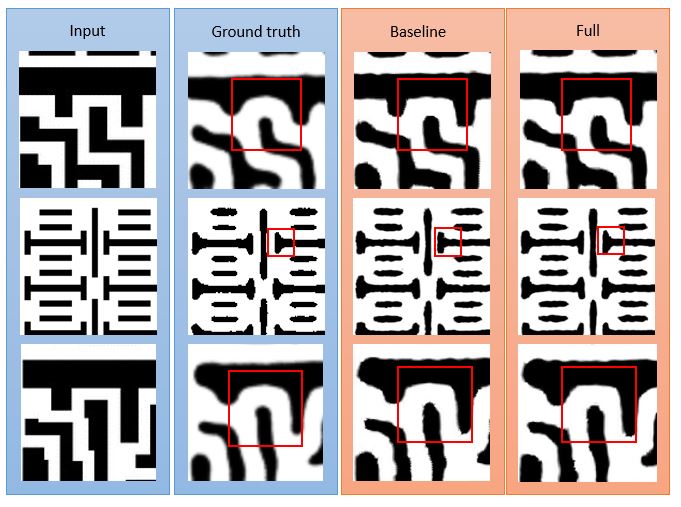}  \vspace{-0.4cm}
\caption{Subject visual quality comparison of LithoNet with and without the total-variation loss $\mathcal{L}_{var}$, where the ``Baseline'' column demonstrates images derived using $\mathcal{L}_{total} - \mathcal{L}_{var}$ and the  ``Full'' column shows predictions synthesized using $\mathcal{L}_{total}$.}
\label{fig:withoutTV}
\end{figure}

The visual effect brought by the total-variation loss $\mathcal{L}_{var}$ is demonstrated in Fig. \ref{fig:withoutTV}, where the ``Baseline'' column demonstrates images derived using $\mathcal{L}_{total} - \mathcal{L}_{var}$, whereas the ``Full'' column shows predictions synthesized  using $\mathcal{L}_{total}$.
This experiment set shows how $\mathcal{L}_{var}$ improves the visual quality of synthetic SEM-styled images. Take regions highlighted by red rectangles in Fig. \ref{fig:withoutTV} for example. Without $\mathcal{L}_{var}$, LithoNet tends to produce straight-line edges and sharp corners, although there are no such patterns on the training images produced by a real IC fabrication process, as shown in  ``Ground truth'' column. By adding $\mathcal{L}_{var}$ to the total loss function, such artifacts can be largely mitigated, thereby more faithfully predicting the shapes of SEM images. 
Finally, note that LithoNet's $\mathcal{L}_{var}$ and $\mathcal{L}_{reg}$ can  be  regarded  as  regularization  terms  to  prevent overfitting.  As  listed  in  the  Tables \ref{table:MStable01}  and  \ref{table:MStable02},  when  LithoNet  was  trained  on  $\mathcal{L}_{total}$, its testing performance is close to that of training data; and, such situation may not hold  for  other  settings,  including  Pix2pix.

\begin{figure}[t]
\centering
\includegraphics[width=0.46\textwidth,keepaspectratio=true]{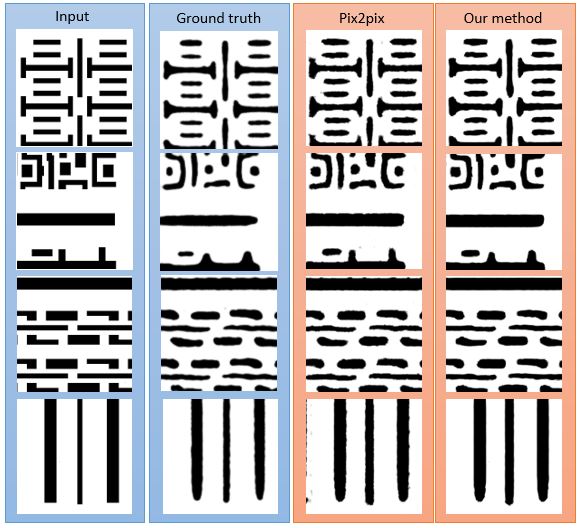} %\\ [-0.4cm]
\vspace{-0.4cm}
\caption{Subjective visual quality comparison between Pix2pix and LithoNet, both trained on UMC dataset \#1.}
\label{fig:lithonet_pix2pix}
\end{figure}

\begin{figure}[!t]
    \centering
    \includegraphics[width=0.50\textwidth,keepaspectratio=true]{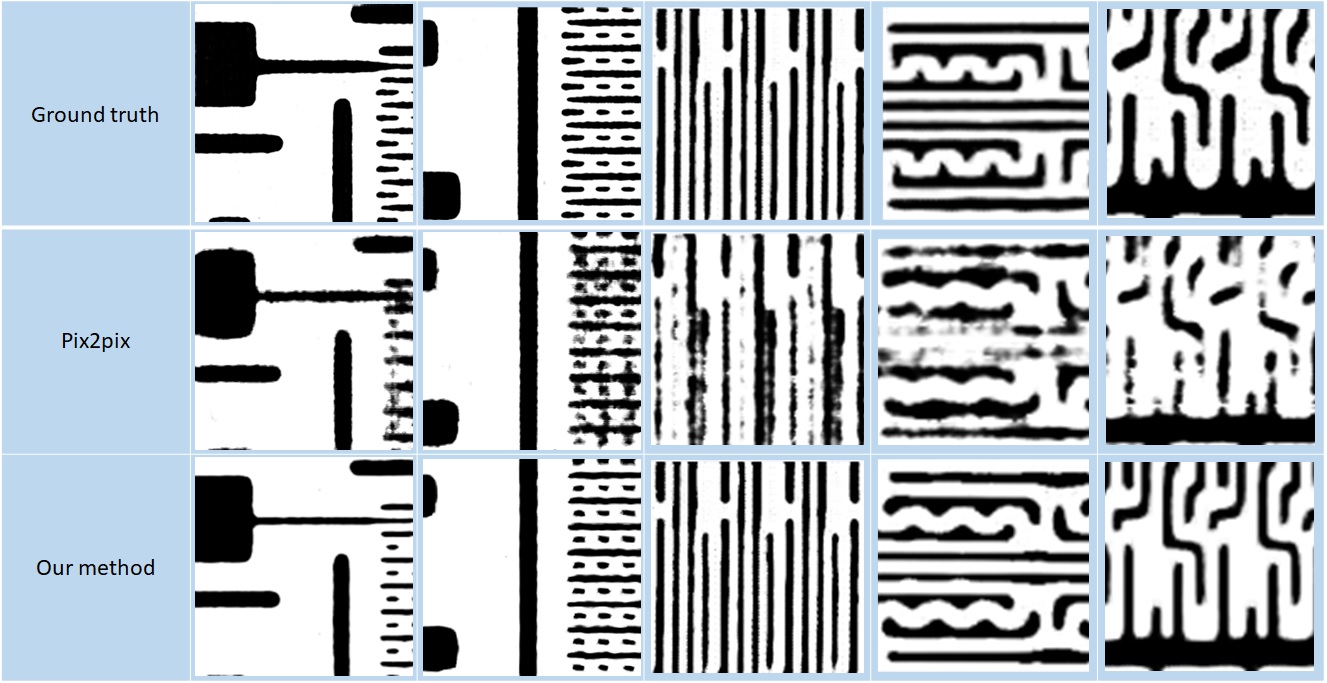} %\par \\ [-0.4cm]
    \vspace{-0.4cm}
    \caption{Subjective visual quality comparison between Pix2pix and LithoNet, both trained on UMC dataset \#1, for some unseen layout patterns of a different observation scale.}
    \label{fig:lithonet_pix2pix_unseenimg}
\end{figure}

\subsubsection{Comparison with Pix2pix}
\label{subsec:405}

As LithoNet is a kind of image-to-image translation scheme, we compare it with Pix2Pix \cite{isola2017image}, a representative GAN-based image-to-image translation method. 
This experiment set was designed for two purposes. 
One is to verify if LithoNet is able to learn special shape correspondences between layout and SEM images, and the other is to check if LithoNet is more advantageous than Pix2Pix in this regard.

As shown in Table \ref{table:MStable01}, Pix2pix achieves slightly higher objective metric values than LithoNet. This situation, however, arises from the fact that these objective metrics mainly reflect the effect of the reconstruction loss term. Nevertheless, compared to Pix2pix, our total loss function described in (\ref{eq:eq01}) contains several additional loss terms, including $\mathcal{L}_{reg}$, $\mathcal{L}_{par}$, and $\mathcal{L}_{smooth}$, which do actually lead to better visual quality as will be explained later. %Consequently, the visual quality of LithoNet is better than Pix2pix, as we will described in next subsection. 

\begin{figure*}[!t]
%\begin{tabular}{ p{185pt}}
\centering
\includegraphics[width=0.9\textwidth,keepaspectratio=true]{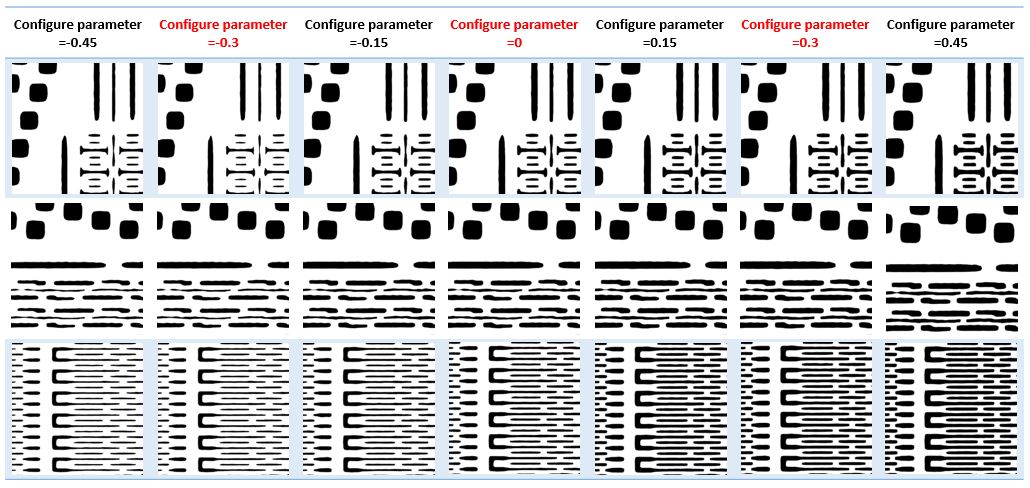} %\par \\  [-0.4cm]
\vspace{-0.4cm}
%\end{tabular}
\caption{Predictions by LithoNet trained on UMC dataset \#2 driven by different configuration parameter values for wafer fabrication. We focus on one configuration parameter which is inversely proportional to the degree of etching: the larger the parameter value, the lower the degree of etching, and the wider the metal lines. Those parameters values used in the training dataset are colored black, whereas those values not used in training are colored red.}
\label{fig:fab_parameters}
\end{figure*}

%--------------------------
%\begin{figure}
%    \centering
%    \includegraphics[width=0.46\textwidth, keepaspectratio=true]{figures/boarderbehavior.jpg} \par \\ [-0.3cm]
%    \caption{Illustrations of border effects. At image borders the shape deformations due to the lithography and etching processes behave differently from those in non-border regions. }
%    \label{fig:MSboarderbehave}
%\end{figure}
%------------------------------

As illustrated in Fig. \ref{fig:lithonet_pix2pix}, Pix2pix produces artifacts like blurred and jiggled contour edges, whereas LithoNet is able to generate clear and smooth ones. Since both Pix2pix and LithoNet utilize $L_1$-norm to guarantee a global shape similarity, this phenomenon would probably be due to the different control strategies over local shapes. Specifically, LithoNet makes use of the total-variation loss, smoothness loss, and regularization loss to control the local deformations, whereas Pix2pix relies on its discriminator architecture, the so-called PatchGAN design that penalizes a structure at the scale of patches, to handle local deformations. 
Consequently, because PatchGAN does not put any penalty on blurred and jiggled edges and learns only to classify if each generated patch looks realistic, such artifacts are reasonable trade-offs of Pix2pix's PatchGAN design.

\begin{table}[t]
\centering
\caption{\small Comparison between LithoNet and Pix2pix, both trained on UMC dataset \#1, for unseen layout patterns of a different scale}
\begin{tabular}{cccccl}
\cline{1-5}
\multicolumn{1}{||c}{Method} & avg & avg & avg & 
\multicolumn{1}{c||}{C2Cdist} & \\ 
\multicolumn{1}{||c}{} & IOU & SSIM & Error & 
\multicolumn{1}{c||}{\small{($\bar{\mu}\pm\bar{\sigma}$)}} & \\ \cline{1-5}
\multicolumn{1}{||c}{Pix2pix}
                         & 0.6587 & 0.6396 & 0.1358 & %0.8179$\pm$2.4623 &
                         \multicolumn{1}{c||}{0.8179$\pm$0.7093}& \\ \cline{1-5} %2.4623} & \\  \cline{1-5}
\multicolumn{1}{||c}{LithoNet}
                         & 0.7107 & 0.6906 & 0.1170 & %0.8010$\pm$2.4570 &
                         \multicolumn{1}{c||}{0.8010$\pm$0.7080}& \\ \cline{1-5} %2.4570} & \\  \cline{1-5}
\end{tabular}
\label{table:MStable03}
\end{table}

\iffalse
\begin{table}[]
\centering
\caption{\small Comparison between LithoNet and Pix2pix, both trained on UMC dataset \#1, for unseen layout patterns of a different scale}
\begin{tabular}{ccccl}
\cline{1-4}
\multicolumn{1}{||c}{Method} & Avg IOU     & Avg SSIM    & \multicolumn{1}{c||}{Avg Error} &  \\ \cline{1-4}
\multicolumn{1}{||c}{Pix2pix}       & 0.6587     & 0.6396      & \multicolumn{1}{c||}{0.1358}    &  \\ \cline{1-4}
\multicolumn{1}{||c}{LithoNet}      & 0.7107     & 0.6906      & \multicolumn{1}{c||}{0.1170}    &  \\ \cline{1-4}
\multicolumn{1}{l}{}               & \multicolumn{1}{l}{} & \multicolumn{1}{l}{} & \multicolumn{1}{l}{}  
\end{tabular}
\label{table:MStable03}
\end{table}
\fi

Fig. \ref{fig:lithonet_pix2pix_unseenimg} compares the prediction results of feeding LithoNet and Pix2pix with test images containing significantly distinct layout patterns from those in the training image set. Moreover, the source dimension of these testing images is much larger than the training data. Therefore, through this experiment we can appraise the reliability and robustness of LithoNet and Pix2pix in mimicking an IC fabrication process when the input layout is a brand new, unseen pattern of a different scale. 
We can observe from  Fig. \ref{fig:lithonet_pix2pix_unseenimg} that, for unseen layout patterns of a different scale,  LithoNet significantly outperforms Pix2pix in terms of the clarity and integrity of shape boundaries, although the predictions of LithoNet still cannot perfectly match the ground-truth for lack of suitable training samples.
Finally, Table \ref{table:MStable03} lists the numerical comparisons between LithoNet and Pix2pix for this case. 

Note that there is still no widely-accepted objective metric to assess the quality of a predicted SEM-styled contour for an IC layout patch with respect to its SEM ground-truth. Some conventional metrics, e.g., IOU and SSIM, measure the similarity globally but ignore local shape discrepancies which may lead to significant impact on IC manufacturability, whereas others, e.g., EPE and EDE \cite{ye2019lithogan}, though designed for shape comparison, still cannot capture local shape discrepancies well. We here leave the problem of developing metrics capable of characterizing both local and global discrepancies and measuring the manufacturablity of a layout pattern  simultaneously and as an open problem for future research.

\subsubsection{Fabrication parameters}
\label{subsec:407}
%Continue configuration parameters editing
Fig. \ref{fig:fab_parameters} compares the predictions by LithoNet trained on UMC dataset \#2 driven by different configuration parameter values for wafer fabrication. 
%We focus on one configuration parameter which is normalized to the range of $[-0.9, 0.9]$ and is inversely proportional to the degree of etching: the larger the parameter value, the lower the degree of etching. 

In this experiment set, we fix the \textit{focus} and adjust the \textit{energy} strength of the scanner in the lithography process to obtain the training samples, and then train LithoNet on them.
We focus on one configuration parameter,
i.e., \textit{energy}, 
normalized to the range of $[-0.9, 0.9]$. 
This parameter is inversely proportional to the degree of etching: the larger the parameter value, the lower the degree of etching. 
This experiment set shows that LithoNet is capable of predicting the width of metal wires by using regression and the discriminator. 

\begin{figure}[t]
\centering
\includegraphics[width=0.40\textwidth,keepaspectratio=true]{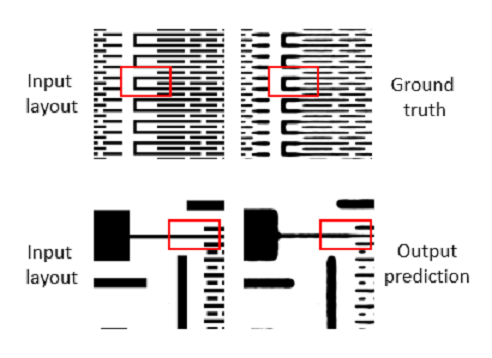} %\par\\ [-0.6cm]
\vspace{-0.6cm}
\caption{Illustrations of interrelationship between the shapes of metal lines and their local neighborhood.}
\label{fig:model_gerality}
\end{figure}
\begin{figure}[!t]
%\center
\centering
%\begin{tabular}{ p{125pt}p{125pt} }%{ p{240pt}p{130pt} }
\includegraphics[width=0.48\textwidth
,keepaspectratio=true]{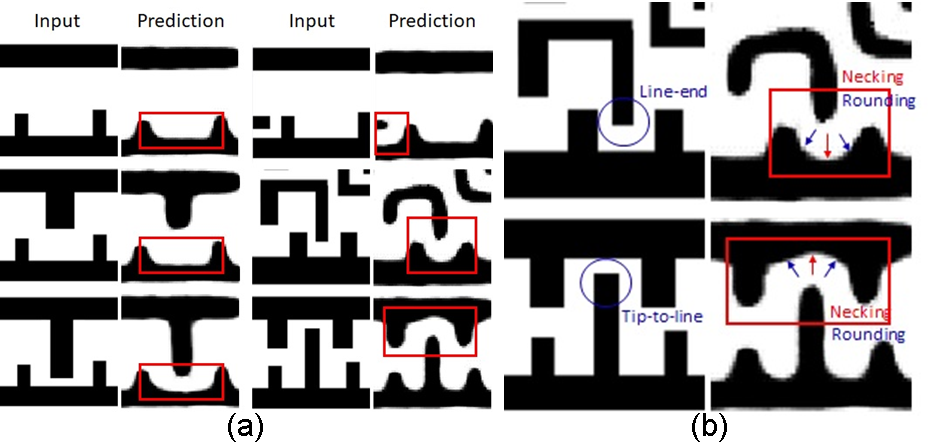} %\par &
%\includegraphics[width=0.22\textwidth%width=0.35\textwidth
%,keepaspectratio=true]{figures/fig14B.jpg} \par  \\%[-0.4cm]
%\vspace{-0.4cm}
%\centerline{(a)} \par &
%\centerline{(b)} \par \\%[-0.5cm]
\vspace{-0.4cm}
%\end{tabular}
\caption{Prediction results of LithoNet: (a) Comparison between a layout and the prediction based on the layout, and (b) conceptual illustration of “Necking” and “Rounding” where the necking effects are highlighted by red boxes and arrows and the rounding effects are indicated by blue arrows.}
\label{fig:neckingrounding}
\end{figure}

Those parameter values used in the training dataset are colored black, and those values not used in training are colored red. This experiment shows that the proposed LithoNet, thank to the regression loss term  $\mathcal{L}_{par}$ described in (\ref{eq:eq06}), does learn the relationship between the line width and the fabrication parameter used to control the degree of etching in the fabrication process. 
Concisely speaking, the larger the parameter is, the wider the metal line should be. 
Hence, our LithoNet model %can not only perform a layout-to-sem image translation task but also 
is able to mimic the fabrication process and generate parameter-dependent prediction results. 
This is an important aspect of LithoNet design, and such design makes LithoNet suitable for semiconductor manufacturing simulations.

%Because the proposed DefNet was trained based on , which was used as 
%
%In this subsection we show the results of using configuration parameters as attribute to train the deformation net. We evaluate our model’s ability to different configuration parameters. Figure \ref{fig:fig08} shows the layout pattern individually transformed with seven intensities of configuration parameters (-4.5, -3, -1.5, 0, 1.5, 3, 4.5). 
%%如果attribute越小的話, 我們預期模擬結果的layout侵蝕越嚴重. 
%The smaller the attribute vector is, the more serious the etching effect. 
%%上句亂寫,要重新看
%From the results we can see the model properly handles these situation and generates simulation results meet fabrication process.

\begin{figure}[!t]
    \centering
    \includegraphics[width=0.46\textwidth,keepaspectratio=true]{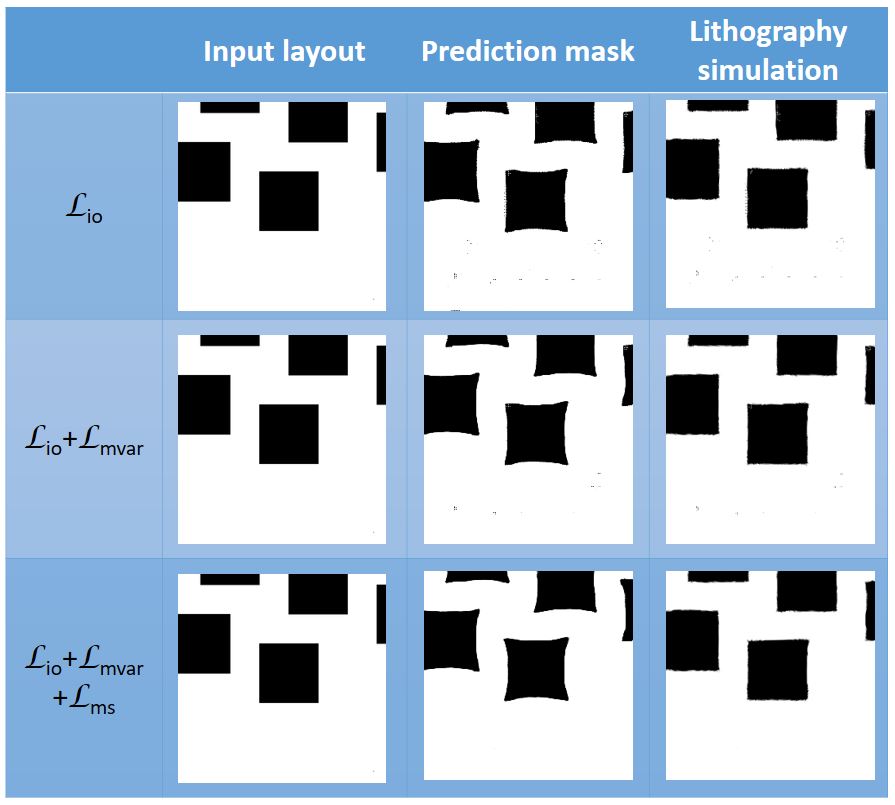} %\par \\ [-0.4cm]
    \vspace{-0.4cm}
    \caption{Illustrations of masks predicted by the mask generator and their lithography simulation outputs. The mean C2Cdist values between layout and lithography simulation of these three cases (from top to bottom) are 10.71, 5.50, and 0.34; and, the standard deviations are 22.73, 16.99, and 0.58.}
    \label{fig:maskloss}
\end{figure}

\begin{figure*}
    \centering
    \includegraphics[width=0.88\textwidth,keepaspectratio=true]{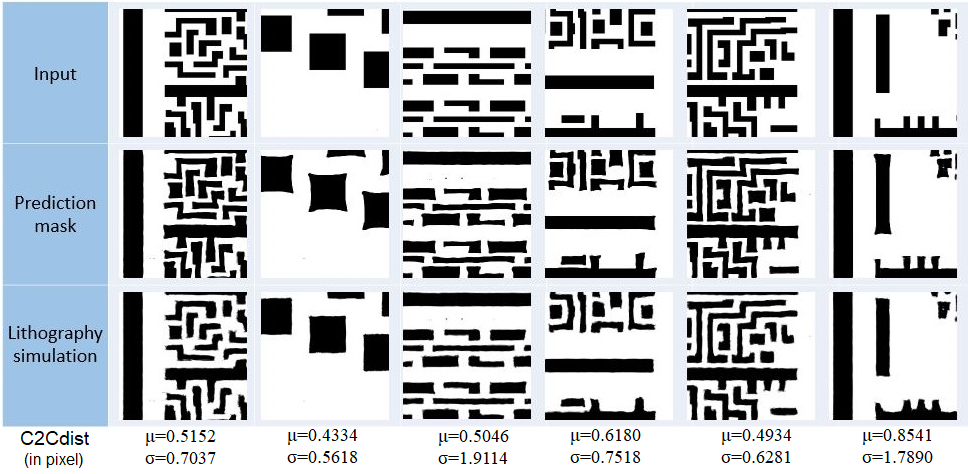}
    \vspace{-0.5cm}
    \caption{%Mask prediction results and their lithography simulation outputs.
    Input layout $\mathcal{S}$, predicted mask $\mathcal{K}$, lithography simulation $\mathcal{J}$, and the \textit{C2Cdist}($\mathcal{S}$, $\mathcal{J}$) value.}
    \label{fig:opcnetprediction}
\end{figure*}

\subsubsection{Model generality}
\label{subsec:408}

Here we examine LithoNet's range of applicability. %by using experiments demonstrated in Figures \ref{fig:fig09} and \ref{fig:fig10}.
%The image pair in the upper row of Figure \ref{fig:fig09} shows the behavior of general semiconductor manufacturing process. As highlighted by the red rectangle, the metal line in an open area is typically wider than its layout design.
%
The image pair in the top row of Fig. \ref{fig:model_gerality} shows that, in an open area, the general fabrication process typically produces a metal line wider than its layout design, as highlighted by the red rectangle. 
The predicted image shown in the bottom row of Fig. \ref{fig:model_gerality} demonstrates that LithoNet learns the shape correspondence between paired training images, so it predicts a wider line in an open area and a narrower one in between two neighboring lines. 
%In addition, in spite of the fact that our prediction does not match perfectly the ground truth, the highlighted regions in Figure \ref{fig:fig10} demonstrate 
%
%In addition, the highlighted regions in Fig. \ref{fig:MSboarderbehave} demonstrate that at image borders the predictions by LithoNet are different from the ground-truths. This is because, at image borders the shape deformations due to the lithography and etching processes behave differently from those in non-border regions, but LithoNet treats them regularly. 
%For example, LithoNet assumes a line reaching a patch border should extend to the adjacent patch rather than shrink from the border. Such border effects can be easily handled by collecting enough training data at image borders, along with  an additional label signifying whether a region is a border one.
Consequently, LithoNet can be expected to forecast fabrication results as long as a large enough amount of training data is given.
%open areas within open areas and to condense lines otherwise.

%In the Figure \ref{fig:fig09}, the region of red rectangle shows the behavior of OPC. At open area, OPC will enlarge the line width like the picture shown. From the simulation result we can see our model learn the OPC behavior from other training pattern and predict in reasonable area. Besides, in Figure \ref{fig:fig10}, the layout border in the red area is exactly the same as the picture boundary, that means in generally we don’t know the length of the line segment. In spite of the fact that our prediction is not as same as ground truth, but it tell us the model predict the result according to the layout image instead of just remember the layout pattern in the training data.

%
% 目前到論文page 30結束,
% Figure 13以下的內容還沒放進來

We also design another experiment to show that LithoNet can learn  the ``necking" and ``rounding" effects that usually occur in IC fabrication, as highlighted by red rectangles in Fig. \ref{fig:neckingrounding}(a) and indicated by the red and blue arrows in Fig. \ref{fig:neckingrounding}(b). Necking is a high-risk pattern caused by either a tip-to-line or a line-end too close to another line on the layout design. As illustrated in Fig. \ref{fig:neckingrounding}(b), such situations may result in a line narrower than designed after fabrication. Hence, this experiment set provides further evidence that a well-trained LithoNet is capable of mimicking the semiconductor lithography and etch processes.

%%%%%%%%%%%%%%%%%%%%%%%%%%%%%%%%%%%%%%%%%%%%%%%%%%%%
%要不要加CycleGAN的比較? 算了先不要加.

%\subsubsection{Comparison with CycleGAN}
%\label{subsec:cycleGAN}
%\textbf{\underline{copy something from our ICCV rebuttal.} Say something about Necking and Rounding.}

%%%%%%%%%%%%%%%%%%%%%%%%%%%%%%%%%%%%%%%%%%%%%%%%%%%%%

\subsection{OPCNet}
\label{subsec:maskgen}

%\subsubsection{Dataset}
%\label{subsec:maskdata}
%We use the dataset as same as the first dataset that be used to train LithoNet. That is, it contains 942 pairs for training, 100 pairs for testing, and each pair has a layout image and its SEM-styled image. Note that we only use the layout images in dataset to train the mask generator unsupervised.

\subsubsection{Impacts of Loss Functions}
\label{subsec:maskdiffloss}

%要寫$\mathcal{L}_{Ksmooth}$只量中間產物，$\mathcal{L}_{Kvar}$量difference，$\mathcal{L}_{consistency}$量基本的L1-norm

%Because OPCNet was trained with the aid of a fixed well-trained LithoNet in an unsupervised manner...

As described in Section \ref{sec:opcnet}, given a layout design pattern $\mathcal{S}$, OPCNet aims to generate a mask $\mathcal{K}$ %\footnote{Ask Pin-Yeng: is this an ``etching"-mask? or a ``photomask".} 
whose lithography simulation result $\mathcal{J}$ predicted by LithoNet is most similar to $\mathcal{S}$. OPCNet is controlled jointly by the IO-consistency loss $\mathcal{L}_{IO}$, the total-variation loss $\mathcal{L}_{Kvar}$, and the mask smoothness loss $\mathcal{L}_{Ksmooth}$. The former two loss terms measure the dissimilarity between $\mathcal{S}$ and $\mathcal{J}$, and the third focuses on the smoothness of $\mathcal{K}$. Here we examine how $\mathcal{L}_{Kvar}$ and $\mathcal{L}_{Ksmooth}$ contribute to the mask prediction task. %etching-mask \textbf{(photomask?)} prediction task. 

Shown in Fig. \ref{fig:maskloss} are three columns of images, each of which corresponds to one loss setting. 
Comparing the mask predicted by using $\mathcal{L}_{IO}$ with that by $\mathcal{L}_{IO}+\mathcal{L}_{Kvar}$, we can find that $\mathcal{L}_{Kvar}$ guarantees the quality of shape contour in the lithography simulation. No matter the $\mathcal{L}_{var}$ of LithoNet or the $\mathcal{L}_{Kvar}$ of OPCNet, such total variation loss 
%measures the sum of magnitudes of high-passed components on pixels around the shape contour, 
accounts for the difference between predicted contours and their ground-truth and focuses on $k$ pixels around the contour pixels. 
%and thus this 
This term helps $\mathcal{L}_{IO}$ guarantee the similarity between the input layout and the lithography simulation and also avoid unexpected artifacts at contours. 
%This is because of that this TV-loss measures the sum of magnitudes of high-passed components on pixels around the shape contour, such loss term can prevent producing unexpected artifacts at contours. \underline{(this paragraph is not yet finished.)}
Finally, comparing the mask predicted by $\mathcal{L}_{IO}+\mathcal{L}_{Kvar}$ with that by $\mathcal{L}_{IO}+\mathcal{L}_{Kvar}+\mathcal{L}_{Ksmooth}$, we find that $\mathcal{L}_{Ksmooth}$ can globally suppress unexpected artifacts on the predicted mask image. The mask prediction derived by $\mathcal{L}_{mask}$ described in (\ref{eq:maskloss}) can thus be artifact-free and smooth.

\subsubsection{Mask Prediction Results}
\label{subsec:predictedmask}

Finally, demonstrated in Fig. \ref{fig:opcnetprediction} are the masks predicted by OPCNet. 
Given a well-trained and accurate lithography simulator LithoNet, Fig. \ref{fig:opcnetprediction} provides evidence that  OPCNet successfully performs the mask optimization task in a self-supervised learning manner without the need of collecting ground-truth OPC-corrected masks. With OPCNet, a layout pattern can be adequately corrected so that the resulting circuit shape best matches the source layout pattern, after an IC-fabrication process.

%\textbf{(this paragraph is not yet finished. Check above three paragraphs.)}

%In the Figure \ref{fig:opcnetprediction} shows the results of mask prediction. Assume our lithography simulator, LithoNet, is accurate enough, that the results in Figure \ref{fig:opcnetprediction} can ensure and prove that our mask generator can successfully do the mask optimization by unsupervised learning. With the mask generator, a layout image can be modified to a mask such that after this mask through IC fabrication process, its wafer image is more similar to the expected layout.
	
%------------------------------------------------------------------------	
	\section{Conclusions}
	\label{sec:conclusion}
	%先亂寫

In this paper we proposed a data-driven framework involving two convolutional neural networks: LithoNet and OPCNet. First, by learning the shape correspondence between paired training images, i.e., IC layout designs and their fabricated IC SEM images, LithoNet can predict the shape deformation field of the layout and then generate a lithography simulation result. Second, with pre-trained LithoNet, OPCNet can learn a mask optimization model without ground-truth OPC-corrected masks based on the proposed input-output consistency loss. Experimental results evidently demonstrate that, in the lithography simulation issue, our method outperforms existing image-to-image translation schemes and the standard compact model-based simulations. In the mask optimization problem, OPCNet can correctly 
predict the mask whose lithography simulation image matches the expected layout. One on-going extension of this work is to establish a scoring system, based on the deformation map or SEM-styled image derived by our method, so that a virtual metrology system for IC circuit layout quality assessment can be developed.
%\textcolor{red}{predict a mask whose lithography simulation image is close to the expected layout}.
%\textcolor{red}{predict a mask for which the lithography simulation image matches the expected layout}.
%that its lithography simulation image is close to the expected layout.

%\textbf{Computational time complexity is less than traditional optics-based methods (ray-tracing??).}
%速度比”傳統光學模擬”?快
%In future work, we can utilize our deformation map to do layout scoring to carry out virtual metrology.
%實現完整的layout模擬跟自動化評估.

	% trigger a \newpage just before the given reference
	% number - used to balance the columns on the last page
	% adjust value as needed - may need to be readjusted if
	% the document is modified later
	%\IEEEtriggeratref{8}
	% The "triggered" command can be changed if desired:
	%\IEEEtriggercmd{\enlargethispage{-5in}}
	
	% references section
	
	% can use a bibliography generated by BibTeX as a .bbl file
	% BibTeX documentation can be easily obtained at:
	% http://mirror.ctan.org/biblio/bibtex/contrib/doc/
	% The IEEEtran BibTeX style support page is at:
	% http://www.michaelshell.org/tex/ieeetran/bibtex/
	\bibliographystyle{IEEEtran}
	\bibliography{gdssem}

% Generated by IEEEtran.bst, version: 1.14 (2015/08/26)
\begin{thebibliography}{10}
\providecommand{\url}[1]{#1}
\csname url@samestyle\endcsname
\providecommand{\newblock}{\relax}
\providecommand{\bibinfo}[2]{#2}
\providecommand{\BIBentrySTDinterwordspacing}{\spaceskip=0pt\relax}
\providecommand{\BIBentryALTinterwordstretchfactor}{4}
\providecommand{\BIBentryALTinterwordspacing}{\spaceskip=\fontdimen2\font plus
\BIBentryALTinterwordstretchfactor\fontdimen3\font minus
  \fontdimen4\font\relax}
\providecommand{\BIBforeignlanguage}[2]{{%
\expandafter\ifx\csname l@#1\endcsname\relax
\typeout{** WARNING: IEEEtran.bst: No hyphenation pattern has been}%
\typeout{** loaded for the language `#1'. Using the pattern for}%
\typeout{** the default language instead.}%
\else
\language=\csname l@#1\endcsname
\fi
#2}}
\providecommand{\BIBdecl}{\relax}
\BIBdecl

\bibitem{watanabe2017accurate}
Y.~Watanabe, T.~Kimura, T.~Matsunawa, and S.~Nojima, ``Accurate lithography
  simulation model based on convolutional neural networks,'' in \emph{Optical
  Microlithography XXX}, vol. 10147, 2017.

\bibitem{ye2019lithogan}
W.~Ye, M.~B. Alawieh, Y.~Lin, and D.~Z. Pan, ``{LithoGAN}: End-to-end
  lithography modeling with generative adversarial networks,'' in
  \emph{ACM/IEEE Design Autom. Conf.}, 2019, pp. 107:1--107:6.

\bibitem{taflove2005computational}
A.~Taflove and S.~C. Hagness, \emph{Computational electrodynamics: the
  finite-difference time-domain method}.\hskip 1em plus 0.5em minus 0.4em\relax
  Artech house, 2005.

\bibitem{lucas1996efficient}
K.~D. Lucas, H.~Tanabe, and A.~J. Strojwas, ``Efficient and rigorous
  three-dimensional model for optical lithography simulation,'' \emph{J.
  Optical Society America: A}, vol.~13, no.~11, pp. 2187--2199, 1996.

\bibitem{otto1994automated}
O.~Otto, J.~Garofalo, K.~K. Low, C.-M. Yuan, R.~Henderson, C.~Pierrat,
  R.~Kostelak, S.~Vaidya, and P.~K. Vasudev, ``Automated optical proximity
  correction: a rules-based approach,'' in \emph{Optical/Laser Microlithography
  VII}, vol. 2197, 1994, pp. 278--294.

\bibitem{hsu2001optical}
T.-J. Hsu, ``Optical proximity correction {(OPC)} method for improving
  lithography process window,'' Feb.~27 2001, uS Patent 6,194,104.

\bibitem{Synopsys}
``{Synopsys}, {Inc}.'' \url{https://www.synopsys.com/}.

\bibitem{aberman2018neural}
K.~Aberman, J.~Liao, M.~Shi, D.~Lischinski, B.~Chen, and D.~Cohen-Or, ``Neural
  best-buddies: sparse cross-domain correspondence,'' \emph{ACM Trans.
  Graphics}, vol.~37, no.~4, p.~69, 2018.

\bibitem{zhou2016learning}
T.~Zhou, P.~Krahenbuhl, M.~Aubry, Q.~Huang, and A.~A. Efros, ``Learning dense
  correspondence via 3d-guided cycle consistency,'' in \emph{Proc. IEEE Conf.
  Comput. Vis. Pattern Recognit.}, 2016, pp. 117--126.

\bibitem{hung2007novel}
M.-H. Hung, T.-H. Lin, F.-T. Cheng, and R.-C. Lin, ``A novel virtual metrology
  scheme for predicting {CVD} thickness in semiconductor manufacturing,''
  \emph{IEEE/ASME Trans. Mechatronics}, vol.~12, no.~3, pp. 308--316, 2007.

\bibitem{susto2015multi}
G.~A. Susto, S.~Pampuri, A.~Schirru, A.~Beghi, and G.~De~Nicolao, ``Multi-step
  virtual metrology for semiconductor manufacturing: A multilevel and
  regularization methods-based approach,'' \emph{Computers \& Operations
  Research}, vol.~53, pp. 328--337, 2015.

\bibitem{poonawala2007mask}
A.~Poonawala and P.~Milanfar, ``Mask design for optical microlithography—an
  inverse imaging problem,'' \emph{IEEE Trans. Image Process.}, vol.~16, no.~3,
  pp. 774--788, 2007.

\bibitem{pan2013design}
D.~Z. Pan, B.~Yu, and J.-R. Gao, ``Design for manufacturing with emerging
  nanolithography,'' \emph{IEEE Trans. Comput.-Aided Design Integr. Circuits
  Syst.}, vol.~32, no.~10, pp. 1453--1472, 2013.

\bibitem{kahng2018reducing}
A.~B. Kahng, ``Reducing time and effort in {IC} implementation: a roadmap of
  challenges and solutions,'' in \emph{Proc. ACM/ESDA/IEEE Design Autom.
  Conf.}, 2018, pp. 1--6.

\bibitem{GANOPC2018yang}
H.~Yang, S.~Li, Y.~Ma, B.~Yu, and E.~F. Young, ``{GAN-OPC}: Mask optimization
  with lithography-guided generative adversarial nets,'' in \emph{Proc.
  ACM/ESDA/IEEE Design Autom. Conf.}, 2018, pp. 1--6.

\bibitem{yu2019deep}
B.-Y. Yu, Y.~Zhong, S.-Y. Fang, and H.-F. Kuo, ``Deep learning-based framework
  for comprehensive mask optimization,'' in \emph{Proc. Asia and South Pacific
  Design Autom. Conf.}, 2019, pp. 311--316.

\bibitem{goodfellow2014generative}
I.~Goodfellow, J.~Pouget-Abadie, M.~Mirza, B.~Xu, D.~Warde-Farley, S.~Ozair,
  A.~Courville, and Y.~Bengio, ``Generative adversarial nets,'' in \emph{Proc.
  Adv. Neural Inf. Process. Syst.}, 2014, pp. 2672--2680.

\bibitem{isola2017image}
P.~Isola, J.-Y. Zhu, T.~Zhou, and A.~A. Efros, ``Image-to-image translation
  with conditional adversarial networks,'' in \emph{Proc. IEEE Conf. Comput.
  Vis. Pattern Recognit.}, 2017, pp. 1125--1134.

\bibitem{wang2018high}
T.-C. Wang, M.-Y. Liu, J.-Y. Zhu, A.~Tao, J.~Kautz, and B.~Catanzaro,
  ``High-resolution image synthesis and semantic manipulation with conditional
  gans,'' in \emph{Proc. IEEE Conf. Comput. Vis. Pattern Recognit.}, 2018, pp.
  8798--8807.

\bibitem{liu2017unsupervised}
M.-Y. Liu, T.~Breuel, and J.~Kautz, ``Unsupervised image-to-image translation
  networks,'' in \emph{Proc. Adv. Neural Inf. Process. Syst.}, 2017, pp.
  700--708.

\bibitem{kingma2013auto}
D.~P. Kingma and M.~Welling, ``Auto-encoding variational bayes,'' \emph{arXiv
  preprint arXiv:1312.6114}, 2013.

\bibitem{zhu2017unpaired}
J.-Y. Zhu, T.~Park, P.~Isola, and A.~A. Efros, ``Unpaired image-to-image
  translation using cycle-consistent adversarial networks,'' in \emph{Proc.
  IEEE Int. Conf. Comput. Vis.}, 2017, pp. 2223--2232.

\bibitem{yi2017dualgan}
Z.~Yi, H.~Zhang, P.~Tan, and M.~Gong, ``{DualGAN}: Unsupervised dual learning
  for image-to-image translation,'' in \emph{Proc. IEEE Int. Conf. Comput.
  Vis.}, 2017, pp. 2849--2857.

\bibitem{bousmalis2017unsupervised}
K.~Bousmalis, N.~Silberman, D.~Dohan, D.~Erhan, and D.~Krishnan, ``Unsupervised
  pixel-level domain adaptation with generative adversarial networks,'' in
  \emph{Proc. IEEE Conf. Comput. Vis. Pattern Recognit.}, 2017, pp. 3722--3731.

\bibitem{huang2018multimodal}
X.~Huang, M.-Y. Liu, S.~Belongie, and J.~Kautz, ``Multimodal unsupervised
  image-to-image translation,'' in \emph{Proc. European Conf. Comput. Vis.},
  2018, pp. 172--189.

\bibitem{gao2014mosaic}
J.-R. Gao, X.~Xu, B.~Yu, and D.~Pan, ``{MOSAIC}: Mask optimizing solution with
  process window aware inverse correction,'' in \emph{Proc. ACM/EDAC/IEEE
  Design Autom. Conf.}, 2014, pp. 52:1--52:6.

\bibitem{gabor2002subresolution}
A.~H. Gabor, J.~A. Bruce, W.~Chu, R.~A. Ferguson, C.~A. Fonseca, R.~L. Gordon,
  K.~R. Jantzen, M.~Khare, M.~A. Lavin, W.-H. Lee \emph{et~al.},
  ``Subresolution assist feature implementation for high-performance logic
  gate-level lithography,'' in \emph{Optical Microlithography XV}, vol. 4691,
  2002, pp. 418--426.

\bibitem{saha2001optimum}
P.~K. Saha and J.~K. Udupa, ``Optimum image thresholding via class uncertainty
  and region homogeneity,'' \emph{IEEE Trans. Pattern Anal. Mach. Intell.},
  vol.~23, no.~7, pp. 689--706, 2001.

\bibitem{otsu1979threshold}
N.~Otsu, ``A threshold selection method from gray-level histograms,''
  \emph{IEEE Trans. Syst., Man, Cybern.}, vol.~9, no.~1, pp. 62--66, 1979.

\bibitem{maninis2018deep}
K.-K. Maninis, S.~Caelles, J.~Pont-Tuset, and L.~Van~Gool, ``Deep extreme cut:
  From extreme points to object segmentation,'' in \emph{Proc. IEEE Conf.
  Comput. Vis. Pattern Recognit.}, 2018.

\bibitem{wang2018deepigeos}
G.~Wang, M.~A. Zuluaga, W.~Li, R.~Pratt, P.~A. Patel, M.~Aertsen, T.~Doel,
  A.~L. David, J.~Deprest, S.~Ourselin \emph{et~al.}, ``{DeepIGeoS}: a deep
  interactive geodesic framework for medical image segmentation,'' \emph{IEEE
  Trans. Pattern Anal. Mach. Intell.}, vol.~41, no.~7, pp. 1559--1572, 2018.

\bibitem{barnich2010vibe}
O.~Barnich and M.~Van~Droogenbroeck, ``{ViBe}: A universal background
  subtraction algorithm for video sequences,'' \emph{IEEE Trans. Image Proc.},
  vol.~20, no.~6, pp. 1709--1724, 2010.

\bibitem{unet2015}
O.~Ronneberger, P.~Fischer, and T.~Brox, ``{U-Net}: Convolutional networks for
  biomedical image segmentation,'' in \emph{Proc. Medical Image Computing
  Computer-Assisted Intervention (MICCAI)}, 2015, pp. 234--241.

\bibitem{jaderberg2015spatial}
M.~Jaderberg, K.~Simonyan, A.~Zisserman \emph{et~al.}, ``Spatial transformer
  networks,'' in \emph{Proc. Adv. Neural Inf. Process. Syst.}, 2015, pp.
  2017--2025.

\bibitem{wang2018discriminative}
C.~Wang, H.~Zheng, Z.~Yu, Z.~Zheng, Z.~Gu, and B.~Zheng, ``Discriminative
  region proposal adversarial networks for high-quality image-to-image
  translation,'' in \emph{Proc. European Conf. Comput. Vis.}, 2018, pp.
  770--785.

\bibitem{rudin1992nonlinear}
L.~Rudin, S.~Osher, and E.~Fatemi, ``Nonlinear total variation based noise
  removal algorithms,'' \emph{Physica D: Nonlinear Phenomena}, vol.~60, no.
  1-4, pp. 259--268, 1992.

\bibitem{maurer2003linear}
C.~Maurer, R.~Qi, and V.~Raghavan, ``A linear time algorithm for computing
  exact euclidean distance transforms of binary images in arbitrary
  dimensions,'' \emph{IEEE Trans. Pattern Anal. Mach. Intell.}, vol.~25, no.~2,
  pp. 265--270, 2003.

\bibitem{SSIM}
Z.~Wang, A.~C. Bovik, H.~R. Sheikh, E.~P. Simoncelli \emph{et~al.}, ``Image
  quality assessment: from error visibility to structural similarity,''
  \emph{IEEE Trans. Image Process.}, vol.~13, no.~4, pp. 600--612, 2004.

\bibitem{C2Cdist}
H.-C. Shao, ``Contour-to-contour distance,''
  \url{https://www.mathworks.com/matlabcentral/fileexchange/75551-contour-to-contour-distance}.

\end{thebibliography}

	\vspace{-0.4in}
\begin{IEEEbiography}
[{\includegraphics[width=1in,height=1.25in,clip,keepaspectratio]{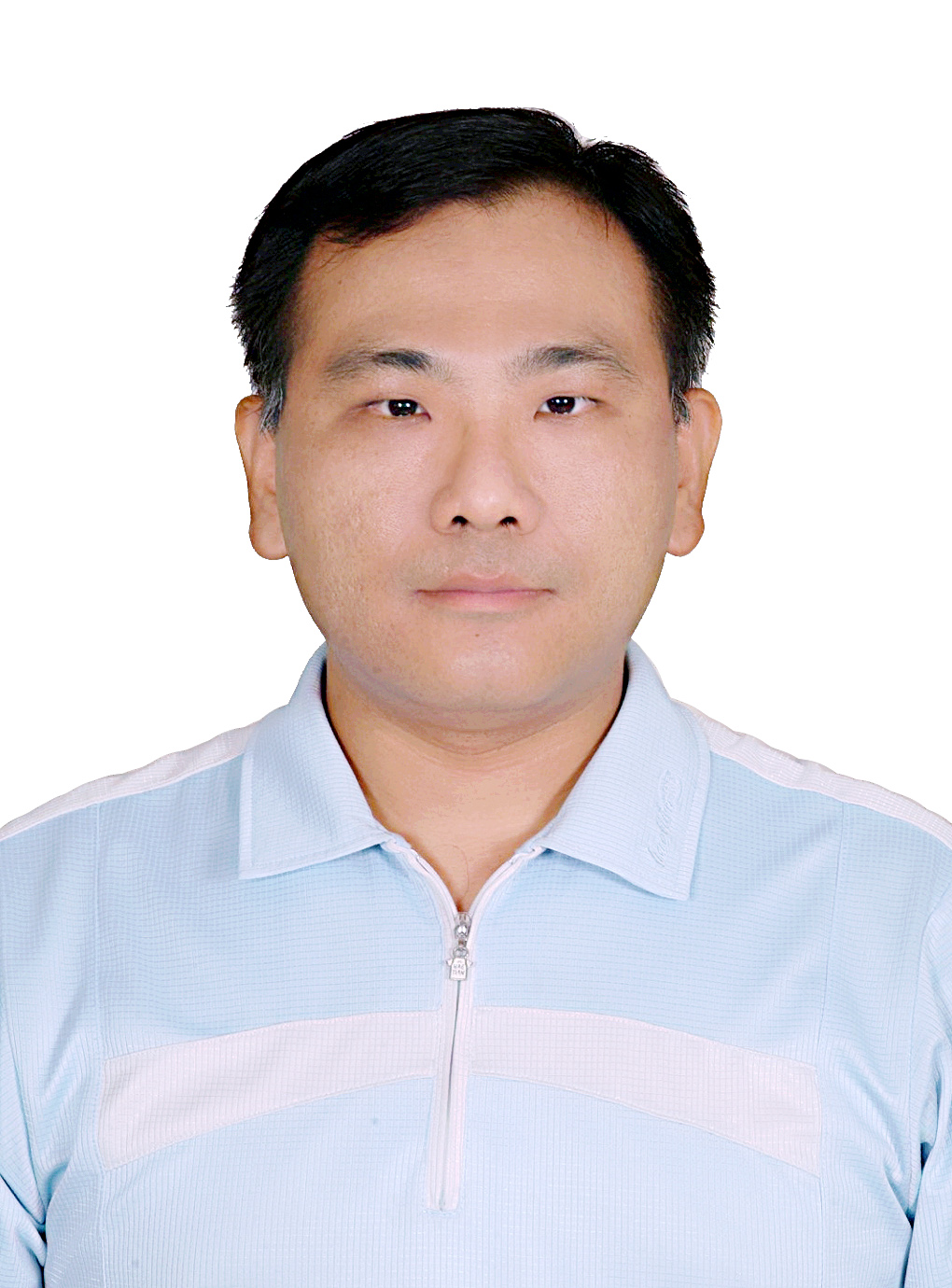}}]
{Hao-Chiang~Shao}
(Member, IEEE) received his Ph.D. degree in electrical engineering from National Tsing Hua University, Taiwan, in 2012. He has been an Assistant Professor with the Dept. Statistics and Information Science, Fu Jen Catholic University, Taiwan, since 2018. During 2012 to 2017, he was a postdoctoral researcher with the Institute of Information Science, Academia Sinica, involved in a series of \textit{Drosophila} brain research projects; in 2017--2018, he was an R\&D engineer with the Computational Intelligence Technology Center, Industrial Technology Research Institute, Taiwan, taking charges of DNN-based automated optical inspection (AOI) projects. His research interests include 2D+Z image atlasing, 3D mesh processing, big industrial image data analysis, and machine learning.
\end{IEEEbiography}

\vspace{-0.4in}
\begin{IEEEbiography}
	[{\includegraphics[width=1in,height=1.25in,clip,keepaspectratio] {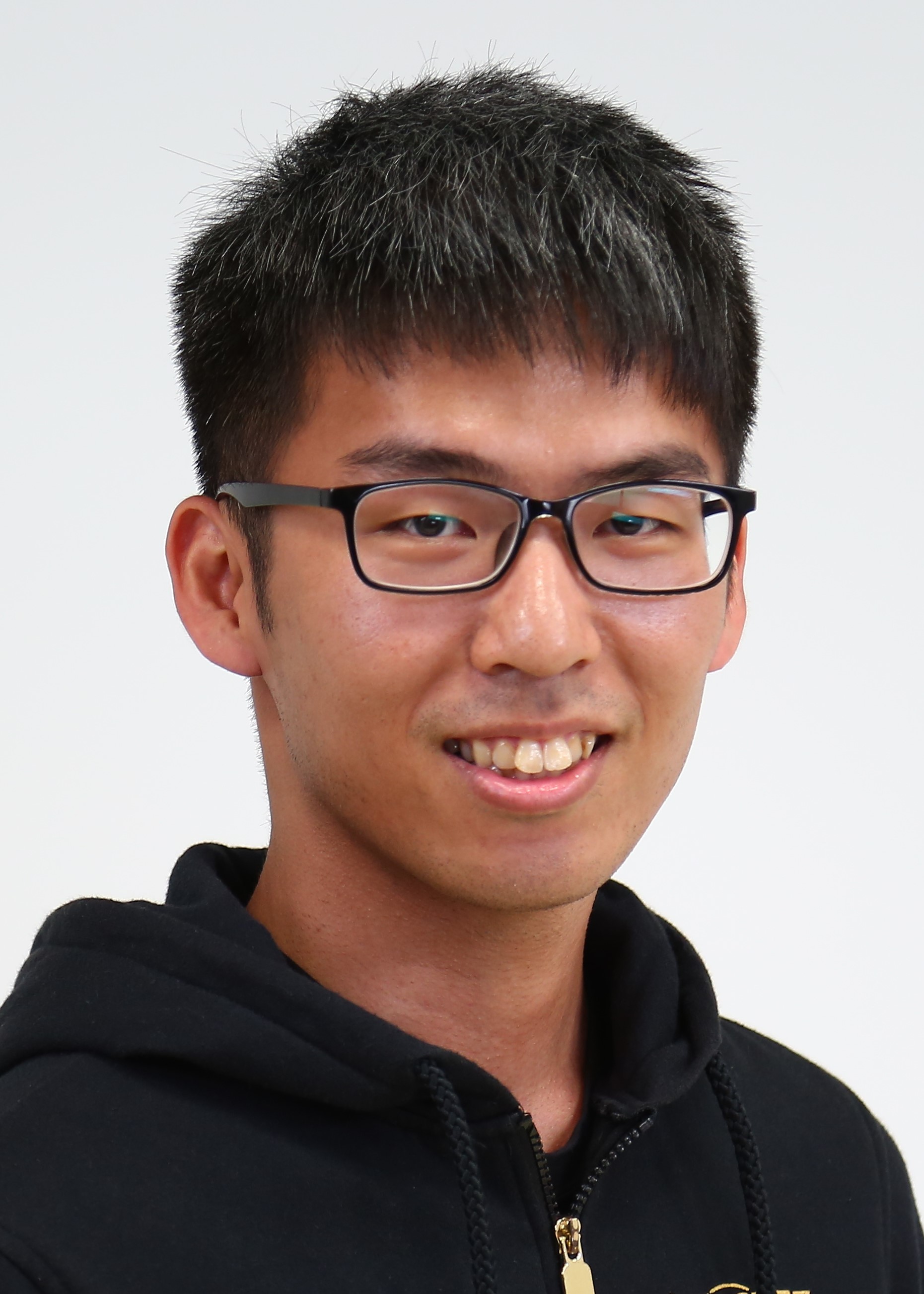}}]
	{Chao-Yi Peng}
	 received his B.S. and M.S. degrees from National Chung Cheng University and National Tsing Hua University, both in Electrical Engineering,  in 2017 and 2019, respectively.  
	He has been working for Altek company as a software engineer since 2019.   His research interests lie in computer vision, machine learning, and visual analytics for IC design for manufacturability.
\end{IEEEbiography}

\vspace{-0.4in}
\begin{IEEEbiography}
	[{\includegraphics[width=1in,height=1.25in,clip,keepaspectratio] {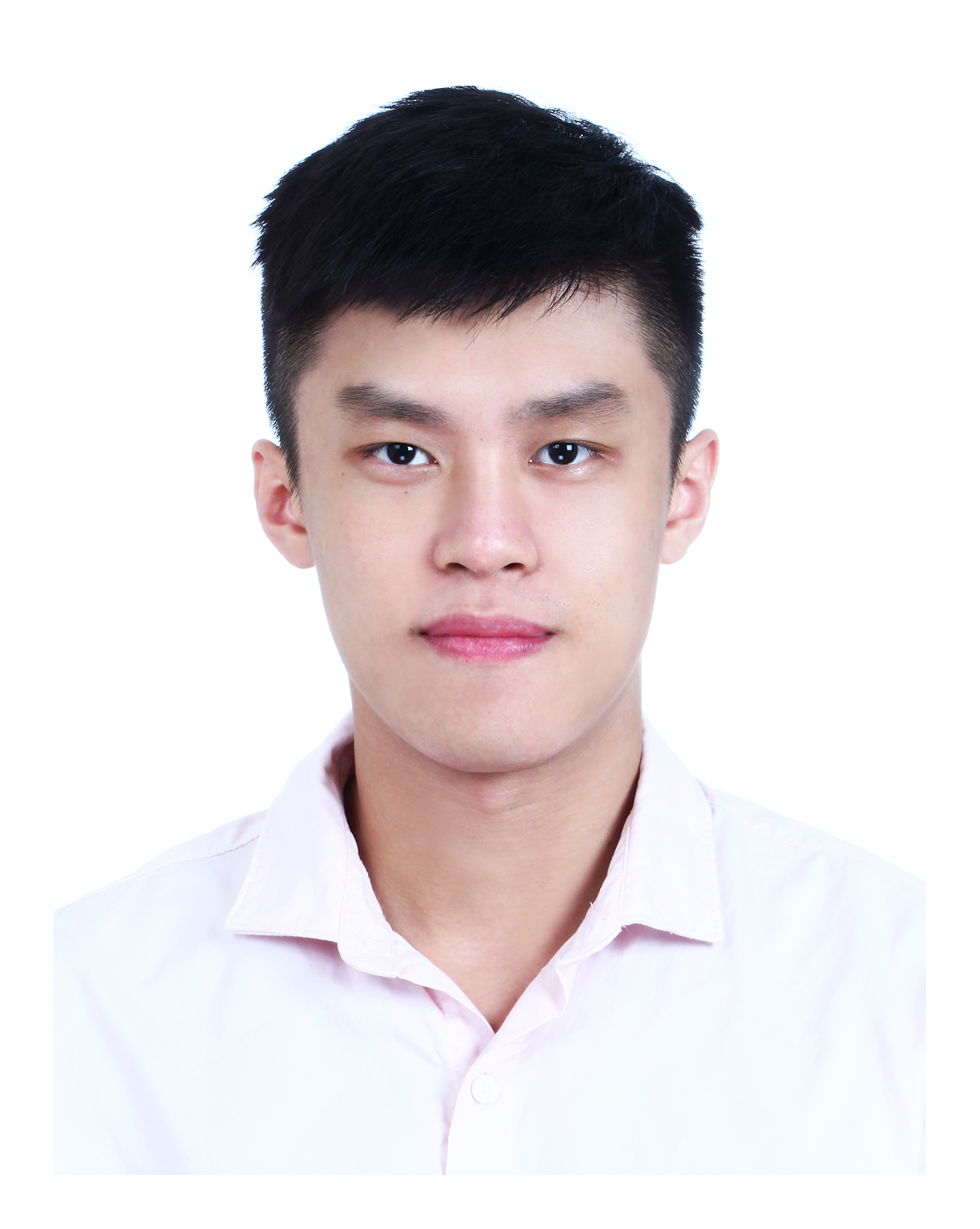}}]
	{Jun-Rei Wu}
	 received his B.S. in Engineering Science and Ocean Engineering from National Taiwan University in 2015 and M.S. degrees in Electrical engineering from National Tsing Hua University  in 2019.  
	He is currently working for HTC VIVE as a software engineer.   His research interests lie in computer vision, machine learning, and visual analytics for IC design for manufacturability.
\end{IEEEbiography}

\vspace{-0.4in}
\begin{IEEEbiography}
	[{\includegraphics[width=1in,height=1.25in,clip,keepaspectratio] {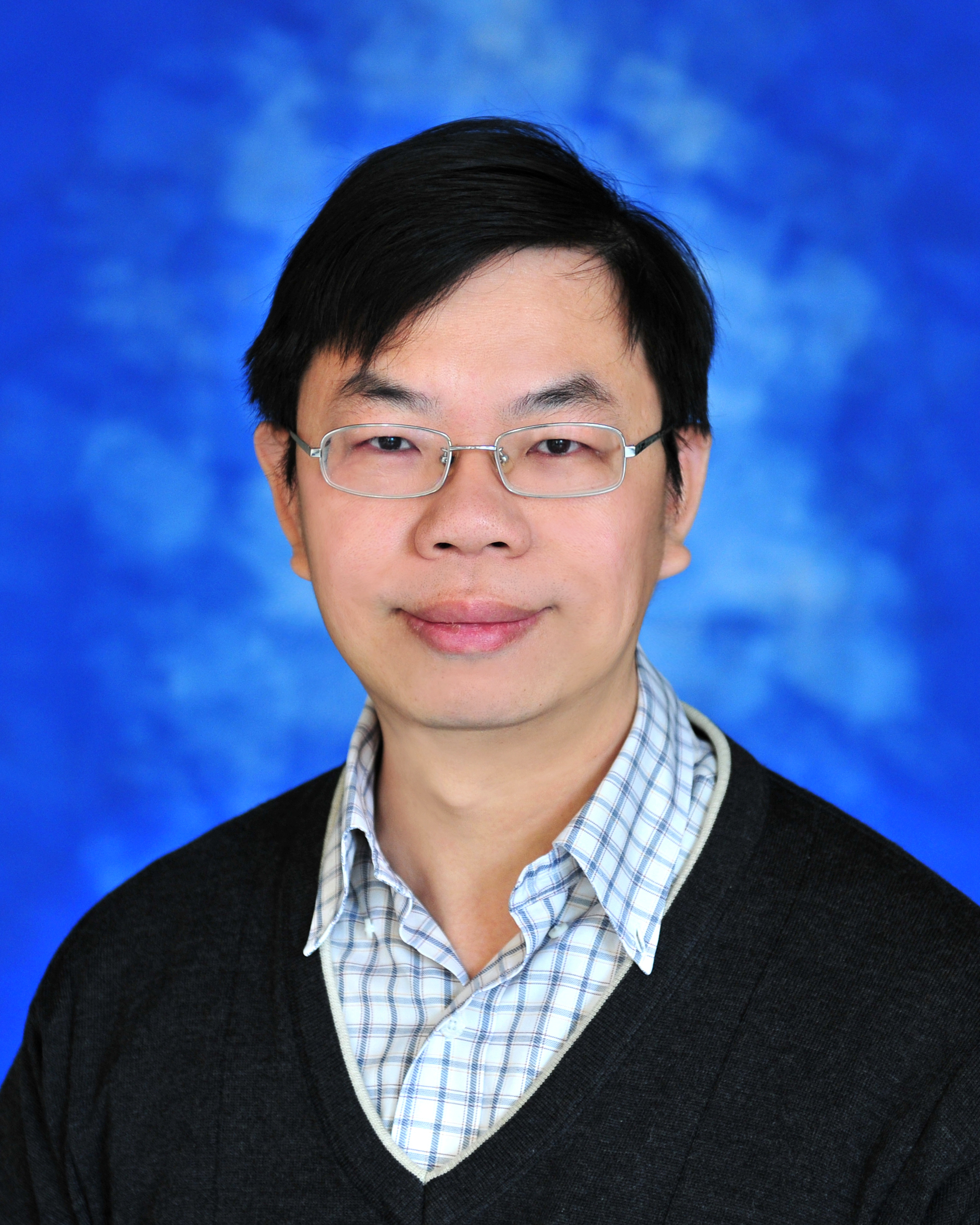}}]
	{Chia-Wen Lin}
	(Fellow, IEEE) received his Ph.D. degree from National Tsing Hua University (NTHU), Taiwan, in 2000.  
	Dr. Lin is currently Professor with the Department of Electrical Engineering and the Institute of Communications Engineering, NTHU.   His research interests include image/video processing, computer vision, and machine learning.  He served as Distinguished Lecturer of IEEE Circuits and Systems Society (2018--2019).   He is Chair of IEEE ICME Steering Committee. He served as TPC Co-Chair of IEEE ICIP 2019 and IEEE ICME 2010, and General Co-Chair of IEEE VCIP 2018. He was a recipient of Outstanding Electrical Engineer Professor Award presented by the Chinese Institute of Electrical Engineering, Taiwan. He received two best paper awards from VCIP 2010 and 2015. He has served as an Associate Editor of \textsc{IEEE Transactions on Image Processing}, \textsc{IEEE Transactions on Circuits and Systems for Video Technology}, \textsc{IEEE Transactions on Multimedia}, and \textsc{IEEE Multimedia}.  He served as a Steering Committee member of \textsc{IEEE Transactions on Multimedia} from 2013 to 2015.
\end{IEEEbiography}

\vspace{-0.4in}
\begin{IEEEbiography}
	[{\includegraphics[width=1in,height=1.25in,clip,keepaspectratio] {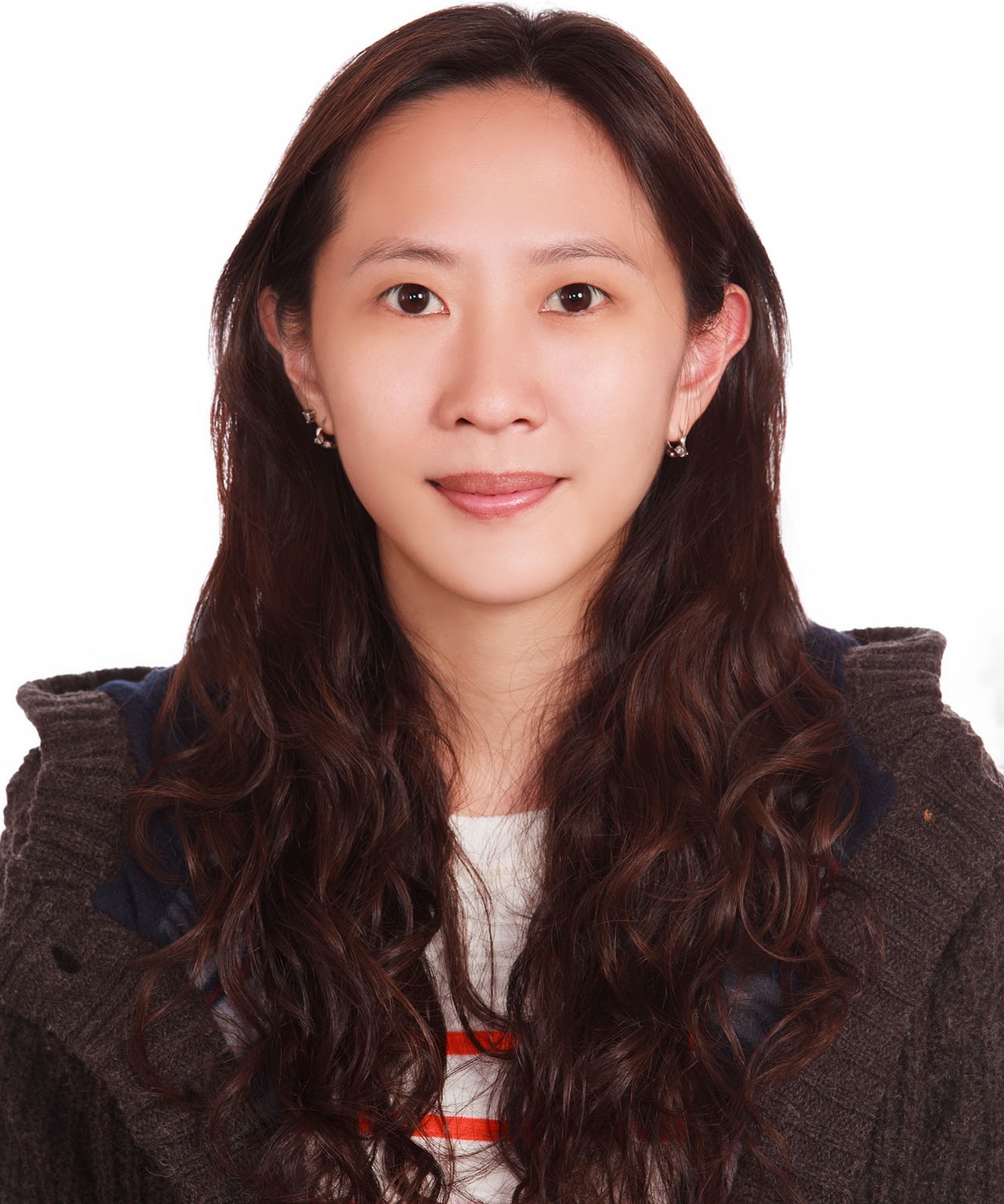}}]
	{Shao-Yun Fang}
	(Member, IEEE) received the B.S. degree in electrical engineering from National Taiwan University (NTU), Taipei, Taiwan, in 2008 and the Ph.D. degree from the Graduate Institute of Electronics Engineering, NTU in 2013. She is currently an Associate Professor of the Department of Electrical Engineering, National Taiwan University of Science and Technology (NTUST), Taipei, Taiwan. Her current research interests focus on physical design and design for manufacturability for integrated circuits. Dr. Fang was the recipient of two Best Paper Awards from the 2016 International Conference on Computer Design and the 2016 International Symposium on VLSI Design, Automation, and Test, and two Best Paper Nominations from the 2012 and 2013 International Symposium on Physical Design.
\end{IEEEbiography}

\vspace{-0.4in}
\begin{IEEEbiography}
	[{\includegraphics[width=1in,height=1.25in,clip,keepaspectratio] {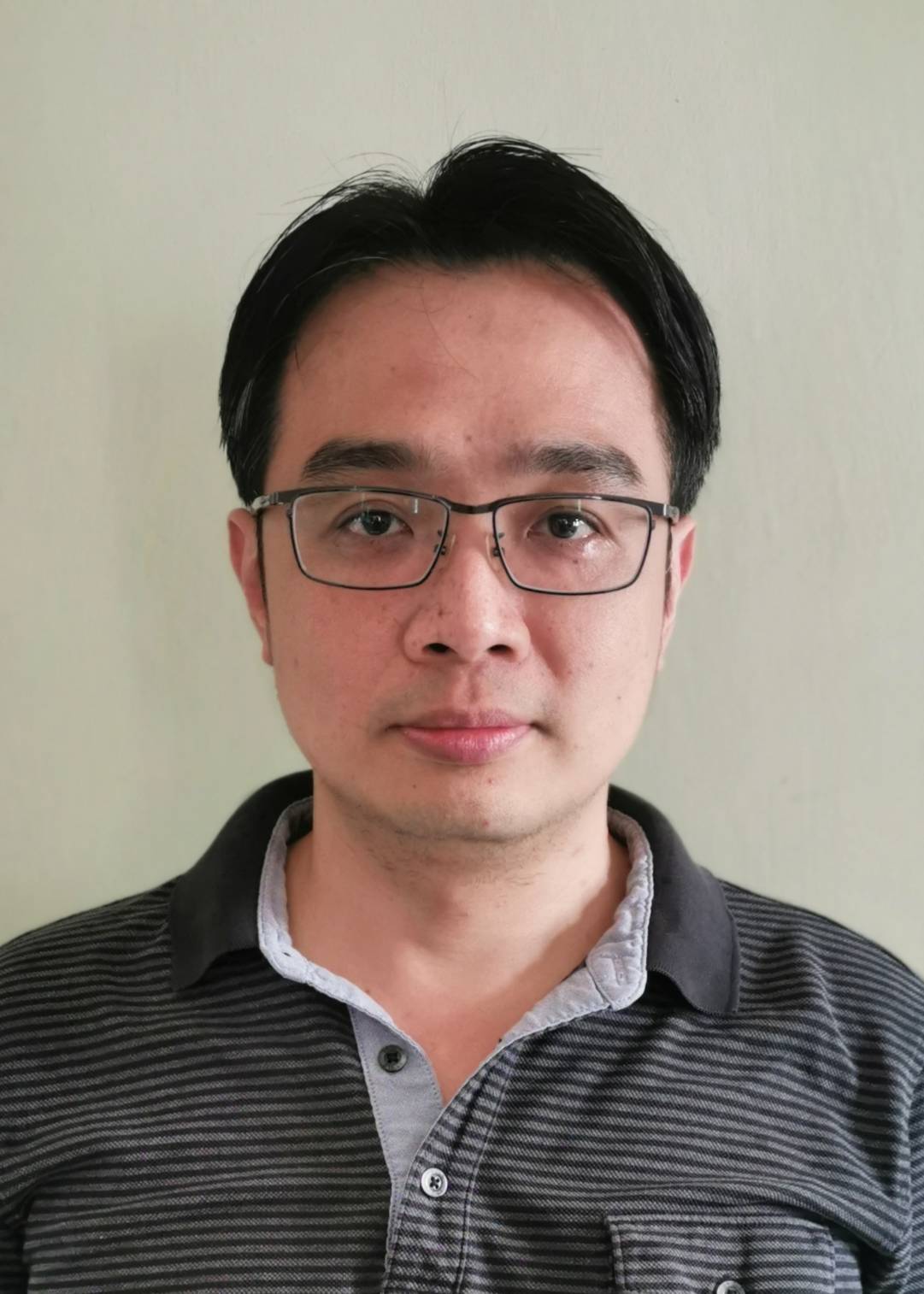}}]
	{Pin-Yian Tsai}
    received his M.S. degree in Physics from National Tsing Hua University (NTHU), Taiwan, in 2008. He is currently a technical manager of the Product Engineering Department in United Microelectronics Corporation (UMC). He led the launch of UMC’s first 14nm product tape out (2017) and is currently working and researching on the field of Design for Manufacturing (DFM). He is now focusing on developing methods for predicting weak patterns in layout manufacturing and automatic optical proximity correction (OPC) to improve the manufacturing yield.
\end{IEEEbiography}

\vspace{-0.4in}
\begin{IEEEbiography}
	[{\includegraphics[width=1in,height=1.25in,clip,keepaspectratio] {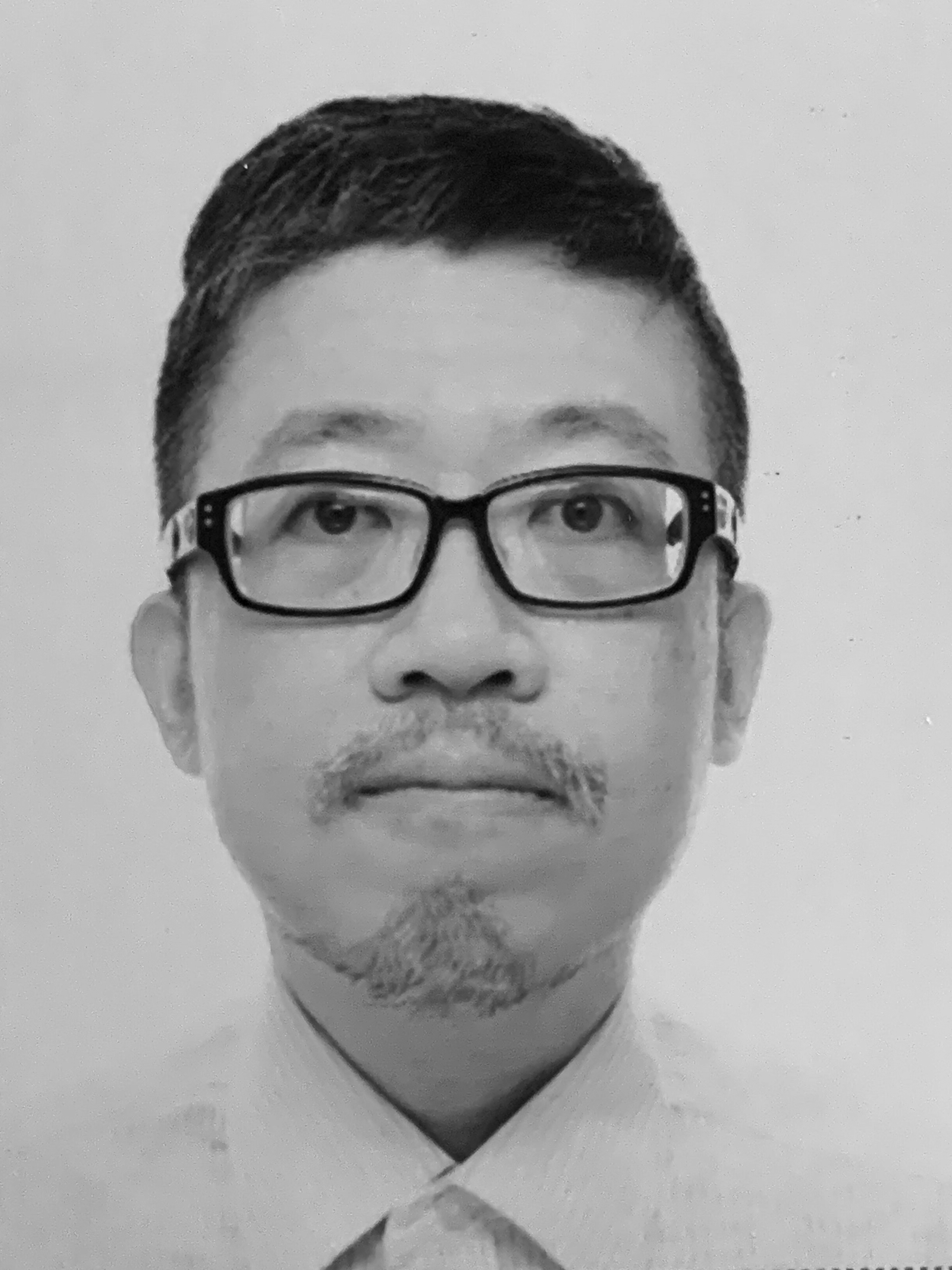}}]
	{Yan-Hsiu Liu}
	received his M.S. degree in Chemistry from National Tsing Hua University (NTHU), Taiwan, in 2002. In 2004, he joined United Microelectronics Corporation (UMC) as a process integration engineer in Hsinchu, Taiwan. He is currently working as a deputy department manager on the development of smart manufacturing and responsible for industry-academia cooperation/collaboration. His research interests include the areas of intelligent manufacturing systems, adaptive parameter estimation, and neural networks.
\end{IEEEbiography}

	% that's all folks
	
%以下幾行先註解掉,
%要產生listoffigure和listoftable時才會需要用到
\iffalse	
\onecolumn

\listoffigures
\newpage

\listoftables
\newpage
\fi

\end{document}